\documentclass[
reprint,
showpacs,
preprintnumbers,
nofootinbib,
amsmath,
amssymb,
aps,
prd,
]{revtex4-1}
\usepackage{enumerate}
\usepackage{color}
\usepackage{graphicx}
\usepackage{dcolumn}
\usepackage{bm}
\usepackage{empheq}
\usepackage{nomencl}
\usepackage{hyperref}
\hypersetup
{
	colorlinks,%
	citecolor=blue,%
	linkcolor=magenta,%
	urlcolor=blue,%
}

\usepackage{subfigure}

\begin{document}
	\newcommand{\red}[1]{\textcolor{red}{#1}}
	\newcommand{\green}[1]{\textcolor{green}{#1}}
	\newcommand{\blue}[1]{\textcolor{blue}{#1}}
	\newcommand{\cyan}[1]{\textcolor{cyan}{#1}}
	\newcommand{\purple}[1]{\textcolor{purple}{#1}}
	\newcommand{\yellowbox}[1]{\colorbox{yellow}{#1}}
	\newcommand{\purplebox}[1]{\colorbox{purple}{#1}}
	\newcommand{\yellow}[1]{\textcolor{yellow!70!red}{#1}}
	\title{Scattering of massless scalar field by charged dilatonic black holes}
	\author{Yang Huang $^1$}\email{sps\_huangy@ujn.edu.cn}
	\author{Hongsheng Zhang $^{1,2}$}\email{sps\_zhanghs@ujn.edu.cn}
	\affiliation{
		$^1$ School of Physics and Technology, University of Jinan, 336, West Road of Nan Xinzhuang, Jinan 250022, Shandong, China\\
		$^2$ State Key Laboratory of Theoretical Physics, Institute of Theoretical Physics, Chinese Academy of Sciences, Beijing, 100190, China}
	
	
	\begin{abstract}
		Wave propagations in the presence of black holes is a significant problem both in theoretical and observational aspects, especially after the discovery of gravitational wave and confirmation of black holes. We study the scattering of massless scalar field by a charged dilatonic black hole in frame of full wave theory. We apply partial wave method to obtain the scattering cross sections of the scalar field, and investigate how the black hole charge affects the scalar scattering cross sections. Furthermore, we investigate the Regge pole approach of the scattering cross section of the dilatonic black hole. We find that in order to obtain results at the same precision, we need more Regge poles as the black hole charge increases. We compare the results in the full wave theory and results in the classical geodesic scattering and the semi-classical glory approximations, and demonstrate the improvements and power of our approach.

	\end{abstract}
	
	\maketitle
	
\section{Introduction}
Black holes are among the most striking predictions in modern physics.
Black hole, as a mathematical solution of the field equation in General Relativity (GR)
or alternative theories of gravity, has been studied many years. In fact, black hole may be the most thoroughly studied object before its discovery. For years the astrophysical black hole is carefully called black hole candidate, or more directly X-ray source. Things changed in 2016 because of the discovery of gravitational wave from binary black hole \cite{Abbott:2016blz}.
After that several gravitational wave events have been reported \cite{GraceDB},
the observation from event horizon telescope (EHT) presents the second clear-cut evidence for the existence of black hole \cite{EHT}.

Among several aspects of investigations of black hole, scattering and absorption from black holes takes a fundamental status, which has been studied more than $50$ years \cite{doi:10.1063/1.1664470}.
This subject attracts increasing attentions due to its relevance with many interesting phenomena,
such as glory, rainbow and superradiant scatterings \cite{futterman_handler_matzner_1988,PhysRevD.31.1869,PhysRevD.22.2331}.
The scattering of scalar ($s=0$), Dirac ($s=1/2$), electromagnetic ($s=1$) and gravitational ($s=2$) waves
by black holes in GR have been thoroughly explored \cite{PhysRevD.100.084025,PhysRevD.10.1059,Glampedakis:2001cx,PhysRevD.18.1798,PhysRevD.74.064005,PhysRevLett.102.231103,Dolan:2008kf,Leite:2019eis,Cotaescu:2014jca}.
These studies have been extended to scattering of plane waves by black holes beyond GR \cite{PhysRevD.92.024012,deOliveira:2018kcq,Huang:2014nka,Gussmann:2016mkp,Lin:2020rvv}
and ultra-compact objects \cite{OuldElHadj:2019kji,Stratton:2019deq,Dolan:2017rtj,Cotaescu:2018etx,Tominaga:1999iy,Tominaga:2000cs}.

In the previous studies of the scattering of waves by black holes, the most frequently approach is based on the partial wave expansion.
This method is natural and straightforward, but it suffers from divergence for small scattering angles
due to the Coulomb characteristic of the potential. 
This problem can be handled by the series reduction method (\cite{PhysRev.95.500}, see also \cite{PhysRevD.74.064005,Leite:2019eis})
or by the Complex Angular Momentum (CAM) techniques, which is first applied to the black hole scattering problem by Andersson and Thylwe \cite{Andersson_1994,Andersson_1994_}.
Recently, the CAM theory has been successfully applied to the scattering of fundamental fields 
by black holes and ultra-compact objects \cite{OuldElHadj:2019kji,Folacci:2019cmc,PhysRevD.100.064009}.

In this paper, we study the scattering of massless scalar field by the Gibbons-Maeda-Garfinkle-Horowitz-Strominger (GMGHS) black hole,
a spherically charged dilatonic solution of the low energy limit of  heterotic string theory in four dimensions,
which was first found by Gibbons and Maeda in \cite{Gibbons:1987ps} and independently obtained by
Garfinkle, Horowitz, and Strominger in \cite{Garfinkle:1990qj} a few years later.

GMGHS black hole has been investigated in theoretical and observational aspects. Particle trajectory around GMGHS black hole was investigated in \cite{Blaga:2014spa}. Late-time evolution of a charged massless scalar field around the GMGHS was studied in \cite{PhysRevD.63.084014}. Hawking radiation of black holes with such a singular horizon was seriously studied in \cite{Zhang:2017zba}. Strong gravitational lensing by GMGHS black hole was explored in \cite{Bhadra:2003zs}, which showed that there are few observational differences between Schwarzchild and  GMGHS black holes for strong lensing. Accretion disks around GMGHS black hole was studied in \cite{Karimov:2018whx}.  
Last but not least, the (in)stability under charged scalar perturbations and the existence of scalar clouds of the GMGHS black hole were studied in \cite{Li:2013jna,Li:2014xxa,Huang:2017whw,Bernard:2017rcw,Bernard:2016wqo,PhysRevD.92.064022,Siahaan:2015xna}.

In Einstein frame, the line element of the GMGHS solution is given by
\begin{equation}
ds^2=-F(r)dt^2+F(r)^{-1}dr^2+r^2G(r)\left(d\vartheta^2+\sin^2\vartheta d\varphi^2\right),
\end{equation}
with
\begin{equation}
	F(r)=1-\frac{2M}{r},\;\text{and}\;G(r)=1-\frac{Q^2}{Mr},
\end{equation}
where $M$ and $Q$ are the mass and charge of the black hole respectively.
The Maxwell field and dilaton field read,
\begin{equation}
	F_M=Q\sin\vartheta d\vartheta\wedge d\varphi,
\end{equation}
and,
\begin{equation}
	e^{-2\phi}=e^{-2\phi_0}\left(1-\frac{Q^2}{Mr}\right),
\end{equation}
respectively.
$\phi_0$ denotes the value of the dilaton $\phi$ at spacelike infinity. $\phi_0=0$ implies an asymptotic flat manifold.
 
The event horizon is located at $r=2M$.
The area of the sphere goes to zero when $r=Q^2/M$ and the surface is singular.
For $Q<Q_{\mathrm{max}}\equiv\sqrt{2}M$, the singularity is enclosed by the event horizon.
In the extremal case $Q=Q_{\mathrm{max}}$, and the singularity coincides with the horizon.
It is convenient to introduce the normalized charge $q=Q/Q_{\mathrm{max}}$.

The reminder of this paper is organized as follows. In Sec. \ref{Sec: geodesic} the geodesic scattering in the GMGHS spacetime is analyzed,
from which we compute the classical scattering cross sections and glory parameters.
In Sec. \ref{Sec: wave src}, we describe the scattering of massless scalar field by the GMGHS black hole,
focusing on computing the scattering cross sections via the partial wave method and the Regge pole approximation.
Our numerical results are presented in Sec. \ref{Sec: results} and we conclude the paper in Sec. \ref{Sec: conclusion}.

\section{Classical and semi-classical scattering}\label{Sec: geodesic}
\subsection{Geodesic scattering}
Under the condition of short wavelength approximation, classical geodesic scattering provides acceptable results of the differential scattering cross section \cite{Collins:1973xf,PhysRevD.92.024012}.
In this subsection, as the first step, we investigate the geodesic scattering in the GMGHS spacetime. 	
The Lagrangian associated to the null geodesics is given by
\begin{equation}\label{Eq: Lagrangian geodesic}
\mathcal{L}=\frac{1}{2}g_{\mu\nu}\dot{x}^\mu\dot{x}^\nu=0,
\end{equation}
where $\dot{x}^\mu=dx^\mu/d\lambda$, and $\lambda$ is an affine parameter.
Since the background spacetime is spherically symmetric,
we shall only consider geodesics on the equatorial plane without loss of generality.
Substituting the GMGHS metric into Eq.(\ref{Eq: Lagrangian geodesic}), we obtain the orbit equation
\begin{subequations}\label{Eq: orbit eq}
	\begin{empheq}{align}
	\left(\frac{du}{d\varphi}\right)^2&=\mathcal{U}(u),\label{Eq: orbit eq1}\\
	\mathcal{U}(u)&=\left(1-ur_-\right)^2\left[\frac{1}{b^2}-\left(\frac{1-ur_+}{1-ur_-}\right)u^2\right],\label{Eq: orbit eq2}
	\end{empheq}
\end{subequations}
where $u=1/r$, $r_+=2M$, $r_-=2Mq^2$, and $b$ is the impact parameter.

From Eq.(\ref{Eq: orbit eq1}) one sees $\mathcal{U}(u)\geq0$ along the orbit.
There is a critical value of the impact parameter $b=b_c$, by which the geodesic may be deflected at any angle.
The critical impact parameter can be obtained by solving equations $\mathcal{U}(u)=0$ and $\mathcal{U}'(u)=0$. The result is:
\begin{equation}\label{Eq: critical impact param}
	\frac{b_c}{M}=\sqrt{\frac{27-q^4-18 q^2+\left(9-q^2\right)\sqrt{\left(1-q^2\right)\left(9-q^2\right)}}{2}}.
\end{equation}
For $b<b_c$, $\mathcal{U}(u)$ is always positive outside the black hole,
and a photon coming from infinity (i.e., $u=0$ initially) will finally be absorbed by the black hole.
In the scattering scenario $b>b_c$, the photon will not be absorbed by the black hole, but will back to infinity
after crossing the turning point $u=u_0$, at which $\mathcal{U}(u_0)=0$. In this case,	
the deflection angle of the geodesic is given by
\begin{equation}\label{Eq: deflection ag}
\Theta(b)=2\int_{0}^{u_0}\frac{du}{\sqrt{\mathcal{U}(u)}}-\pi.
\end{equation}
We apply the numerical integration to obtain the deflection angle.
In the weak-field limit ($b\gg M$), it is possible to find the analytic expression of $\Theta$, i.e.,
\begin{equation}\label{Eq: deflection ag weak-field}
	\Theta\approx\frac{4M}{b}+\frac{3\pi}{4}\left(5-4q^2-\frac{4}{3}q^4\right)\frac{M^2}{b^2}.
\end{equation}
Note that this equation is slightly different from that of the Reissner-Nordstr\"{o}m black hole \cite{PhysRevD.79.064022}, and the first term is the term of a typical strong lensing.

\begin{figure}
	\centering
	\includegraphics[width=0.42\textwidth,height=0.29\textwidth]{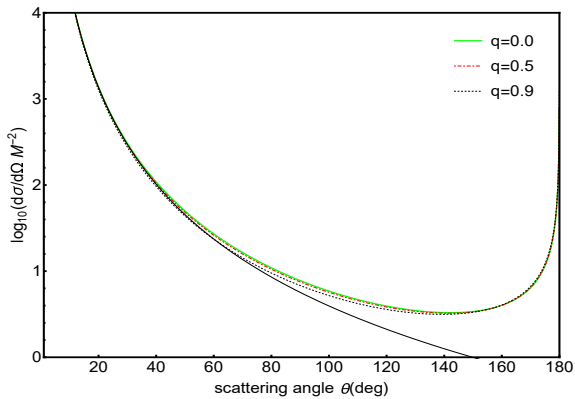}			
	\caption{Classical scattering cross section of GMGHS black holes, with $q=0,\;0.5$ and $0.9$.
		The black solid line is given by Eq.(\ref{Eq: classical src small th}), with $q=0$.}
	\label{Fig: classical src}
\end{figure}

Given the relation between $b$ and the deflection angle, the classical differential scattering cross section is presented by
\begin{equation}\label{Eq: classical src}
	\frac{d\sigma}{d\Omega}=\frac{b}{\sin\theta}\bigg|\frac{db}{d\theta}\bigg|.
\end{equation}
The effects of $q$ on the classical scattering cross section is exhibited in Fig. \ref{Fig: classical src}.
This figure shows that for moderate angles, the classical scattering cross section decreases as the black hole charge $q$ increases.
But this effect is not obvious for small values of $\theta$. The interpretation of this result is as follows:
combining Eqs.(\ref{Eq: deflection ag weak-field}) and (\ref{Eq: classical src}), we obtain
\begin{equation}\label{Eq: classical src small th}
\frac{d\sigma}{d\Omega}\approx\frac{16M^2}{\theta^4}+\frac{3\pi M^2}{4\theta^3}\left(5-4q^2-\frac{4}{3}q^4\right).
\end{equation}
This equation shows clearly that the black hole charge $q$ does not contribute to the dominant term of the classical scattering cross section
for small values of $\theta$.

\subsection{Glory scattering}
The glory approximation of the scalar scattering cross sections is \cite{PhysRevD.31.1869}
\begin{equation}\label{Eq: glory}
	\frac{d\sigma}{d\Omega}=2\pi\omega b^2_g\bigg|\frac{db}{d\theta}\bigg|_{\theta=\pi}J^2_{0}(\omega b_g\sin\theta),
\end{equation}
where $J_{0}(x)$ is the Bessel function of the first kind, and $b_g$ is the impact parameter of backscattered geodesics ($\theta=\pi$).
There are multiple values of $b_g$ corresponding to the multiple values of the deflection angle, i.e., $\theta=\pi+2n\pi$
(with $n=0,1,2,\cdots$ the number of times that the null geodesic rotates around the black hole),
but the major contribution to the differential scattering cross section comes from the $n=0$ case \cite{PhysRevD.92.024012}.
Although the semi-classical glory approximation (\ref{Eq: glory}) is expected to be valid at high frequencies ($\omega M\gg1$),
it is still in good agreement with the numerical results for intermediate frequencies, i.e., $\omega M\sim1$ \cite{PhysRevD.79.064022}.

\begin{figure}
	\centering	
	\includegraphics[width=0.42\textwidth,height=0.29\textwidth]{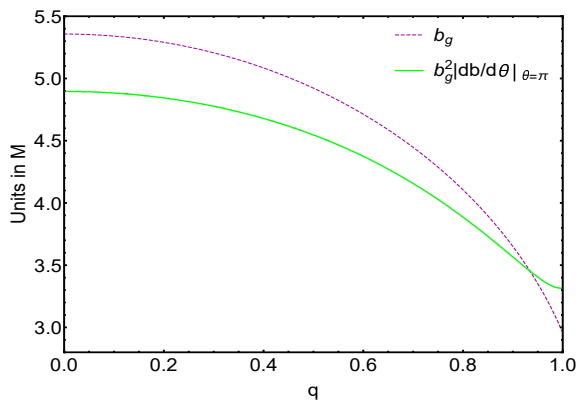}			
	\caption{Glory scattering parameters as functions of $q$.}
	\label{Fig: glory param}
\end{figure}

According to Eq.(\ref{Eq: glory}), we need to determine $b_g$ and $|db/d\theta|_{\theta=\pi}$ in order to obtain the glory scattering cross section.
Given a $q$, we obtain $b_g$ by numerically solving the equation $\Theta(b)=\pi$, where $\Theta(b)$ is given in Eq.(\ref{Eq: deflection ag}).
Then, it is straightforward to obtain $|db/d\theta|_{\theta=\pi}$ via the finite difference formula.
Numerical results of $b_g$ and $b^2_g|db/d\theta|_{\theta=\pi}$ are presented in Fig.\ref{Fig: glory param}.
As is shown in the plot, both $b_g$ and $b^2_g|db/d\theta|_{\theta=\pi}$ decrease monotonically as the black hole charge $q$ increases.
These results indicate that with the increase of $q$,
(i) interference fringes get wider, since the interference fringe width is inversely proportional to $b_g$;
(ii) the backscattered flux intensity decreases.

\section{Wave scattering}\label{Sec: wave src}
\subsection{Massless scalar field in the GMGHS spacetime}
We consider a massless scalar field propagating in a GMGHS spacetime, which obeys the Klein-Gordon equation
\begin{equation}\label{Eq: KG eq}
\nabla_\mu\nabla^\mu\Phi=0.
\end{equation}
The background spacetime is stationary and spherically symmetric. Thus, we write $\Phi=R_{\omega l}(r)Y_{lm}(\vartheta,\varphi)e^{-i\omega t}$,
where $Y_{lm}(\vartheta,\varphi)$ are the scalar spherical harmonics. The radial function $R_{\omega l}(r)$ obeys the radial equation
\begin{equation}\label{Eq: the radial eq}
\Delta\frac{d}{dr}\left(\Delta\frac{dR_{\omega l}}{dr}\right)+\left[G(r)^2\omega^2r^4-\Delta l(l+1)\right]R_{\omega l}=0,
\end{equation}
where we have introduced a new function $\Delta=(r-r_+)(r-r_-)$. By introducing $\psi_{\omega l}(r)=R_{\omega l}(r)/r$,
we can rewrite the radial equation in the following form
\begin{equation}\label{Eq: the radial eq psi}
\frac{d^2}{dr^2_*}\psi_{\omega l}(r)+\left[G(r)^2\omega^2-V_l(r)\right]\psi_{\omega l}(r)=0,
\end{equation}
where the tortoise coordinate is defined by $dr_*=f(r)^{-1}dr$, and $f(r)=\Delta/r^2$, while the potential is given by
\begin{equation}
V_l(r)=f(r)\left[\frac{f'(r)}{r}+\frac{l(l+1)}{r^2}\right].
\end{equation}

At the event horizon $r=2M$, the purely ingoing wave solution of the radial equation is
\begin{equation}\label{Eq: bcs horizon}
\psi(r)\sim e^{-i\kappa\omega r_*},
\end{equation}
where $\kappa=1-Q^2/2M^2$. At spatial infinity $r\rightarrow\infty$, the radial solution behaves as
\begin{equation}\label{Eq: bcs infty}
\psi(r)\sim A^{(-)}_{l}(\omega)e^{-i\omega r_*}+A^{(+)}_{l}(\omega)e^{i\omega r_*}.
\end{equation}
Here the coefficients $A^{(-)}_{l}(\omega)$ and $A^{(+)}_{l}(\omega)$ are complex amplitudes of
the ingoing and outgoing waves, respectively. Given these amplitudes,
the S-matrix elements $S_{l}(\omega)$ is given by
\begin{equation}
S_{l}(\omega)=e^{i(l+1)\pi}\frac{A^{(+)}_{l}(\omega)}{A^{(-)}_{l}(\omega)}.
\end{equation}
There are two types of poles of $S_l(\omega)$: (i) for $l\in\mathbb{N}$, the poles of $S_{l}(\omega)$ in the complex plane of $\omega$ give the quasinormal frequencies,
and they are associated to the dynamics of scalar perturbations around the GMGHS black hole;
(ii) Regge poles are the poles of\footnote{Here $S_{\lambda-1/2}(\omega)$ denotes the analytic extension of $S_{l}(\omega)$.} $S_{\lambda-1/2}(\omega)$
in the first and third quadrants of the complex plane of $\lambda=l+1/2$ for $\omega\in\mathbb{R}$,
and they can be expressed as $\lambda=\lambda_{n}(\omega)$ with $n=1,2,3,\cdots$.

\subsection{Scattering cross section}
The differential scattering cross section is given by
\begin{equation}\label{Eq: scalar scattering cross section}
	\frac{d\sigma}{d\Omega}=|f(\omega,\theta)|^2.
\end{equation}
where $f(\omega,\theta)$ the scattering amplitude, which is expressed by the following partial wave series
\begin{equation}\label{Eq: partial wave expansion}
	f(\omega,\theta)=\frac{1}{2i\omega}\sum_{l=0}^{\infty}(2l+1)\left[S_{l}(\omega)-1\right]P_l(\cos\theta).
\end{equation}
 
In order to derive the scattering cross section via the partial wave series,
we need to compute $S_{l}(\omega)$ by solving the radial equation (\ref{Eq: the radial eq psi}) (or equivalently Eq.(\ref{Eq: the radial eq}))
with respect to the boundary conditions given in Eqs.(\ref{Eq: bcs horizon}) and (\ref{Eq: bcs infty}).
We apply the Runge-Kutta method to solve this problem numerically.
Note that the sum in Eq.(\ref{Eq: partial wave expansion}) does not convergent very quickly for small values of $\theta$.
Thus, we employ the method developed in \cite{PhysRev.95.500}, and first applied to the black hole scattering problem in \cite{PhysRevD.74.064005},
to improve the convergence of the sum.

\subsection{Regge pole approximation}
According to the CAM theory, the scattering amplitude can be split into a background integral and a sum over Regge poles \cite{Andersson_1994_,Andersson_1994}
\begin{equation}\label{Eq: Regge pole approx}
	f_{\text{P}}(\omega,\theta)=-\frac{i\pi}{\omega}\sum_{n=1}^{\infty}\frac{\lambda_{n}(\omega)r_n(\omega)}{\cos\left[\pi\lambda_{n}(\omega)\right]}P_{\lambda_{n}(\omega)-1/2}(-\cos\theta),
\end{equation}
where $P_{\lambda-1/2}(x)$ denotes the analytic extension of the Legendre polynomials $P_{l}(x)$,
and $r_{n}(\omega)$ are the residues of the matrix $S_{\lambda-1/2}(\omega)$ at $\lambda=\lambda_{n}(\omega)$,
they are given by
\begin{equation}
	r_{n}(\omega)=e^{i\pi\left[\lambda_{n}(\omega)-1/2\right]}\left[\frac{A^{(+)}_{\lambda-1/2}(\omega)}{\frac{d}{d\lambda}A^{(-)}_{\lambda-1/2}(\omega)}\right]_{\lambda=\lambda_{n}(\omega)},
\end{equation}
where the complex amplitudes $A^{(-)}_{\lambda-1/2}(\omega)$ and $A^{(-)}_{\lambda-1/2}(\omega)$ are defined from the analytic extension of Eq.(\ref{Eq: bcs infty}).		
It was shown that the contribution from the background integral is negligible for high frequencies \cite{Folacci:2019cmc}. %
Hence, the sum over Regge poles (\ref{Eq: Regge pole approx}) provides a good approximation of the scalar differential cross section.

\begin{figure}
	\centering	
	\includegraphics[width=0.42\textwidth,height=0.29\textwidth]{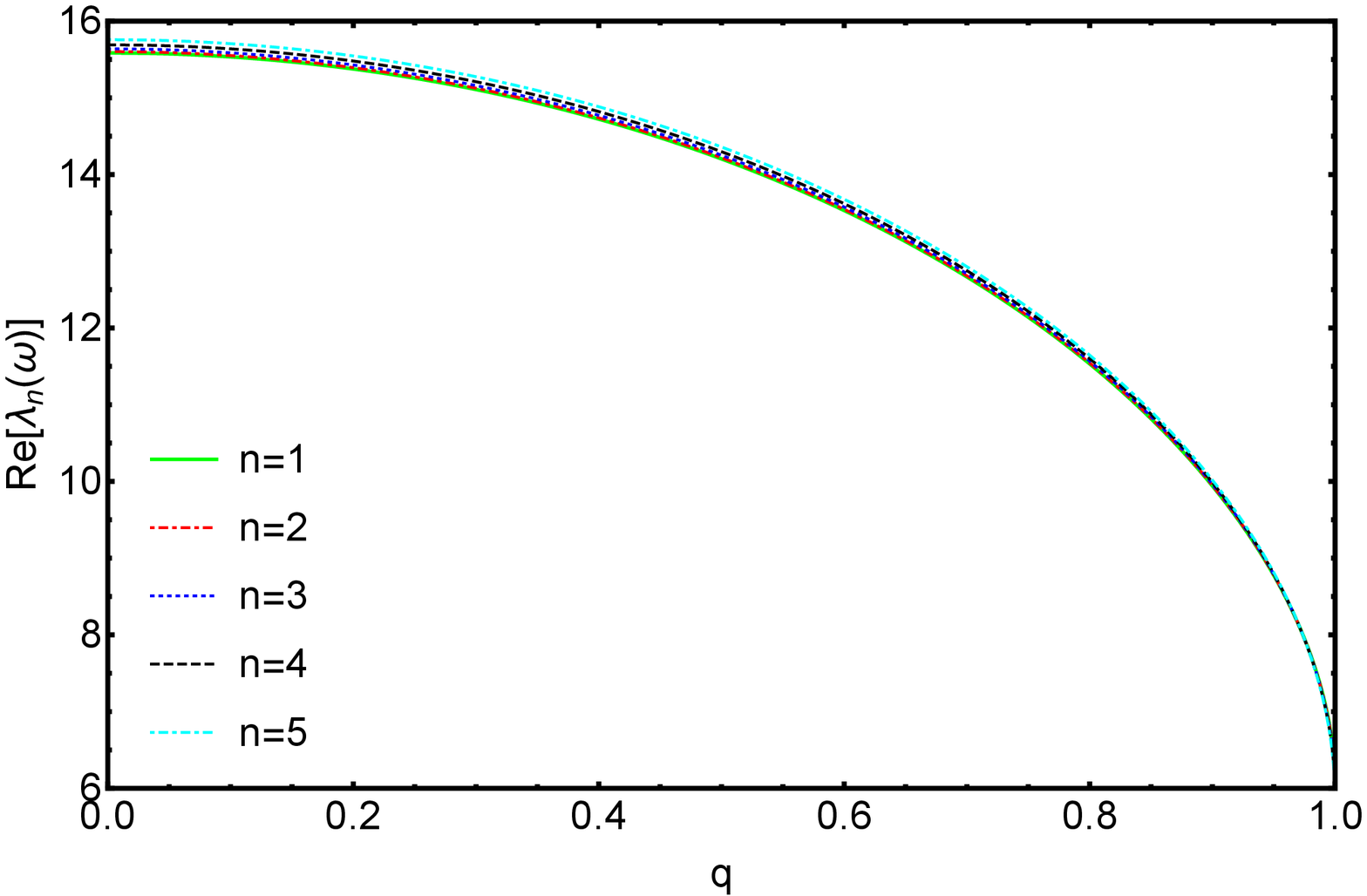}	
	\includegraphics[width=0.42\textwidth,height=0.29\textwidth]{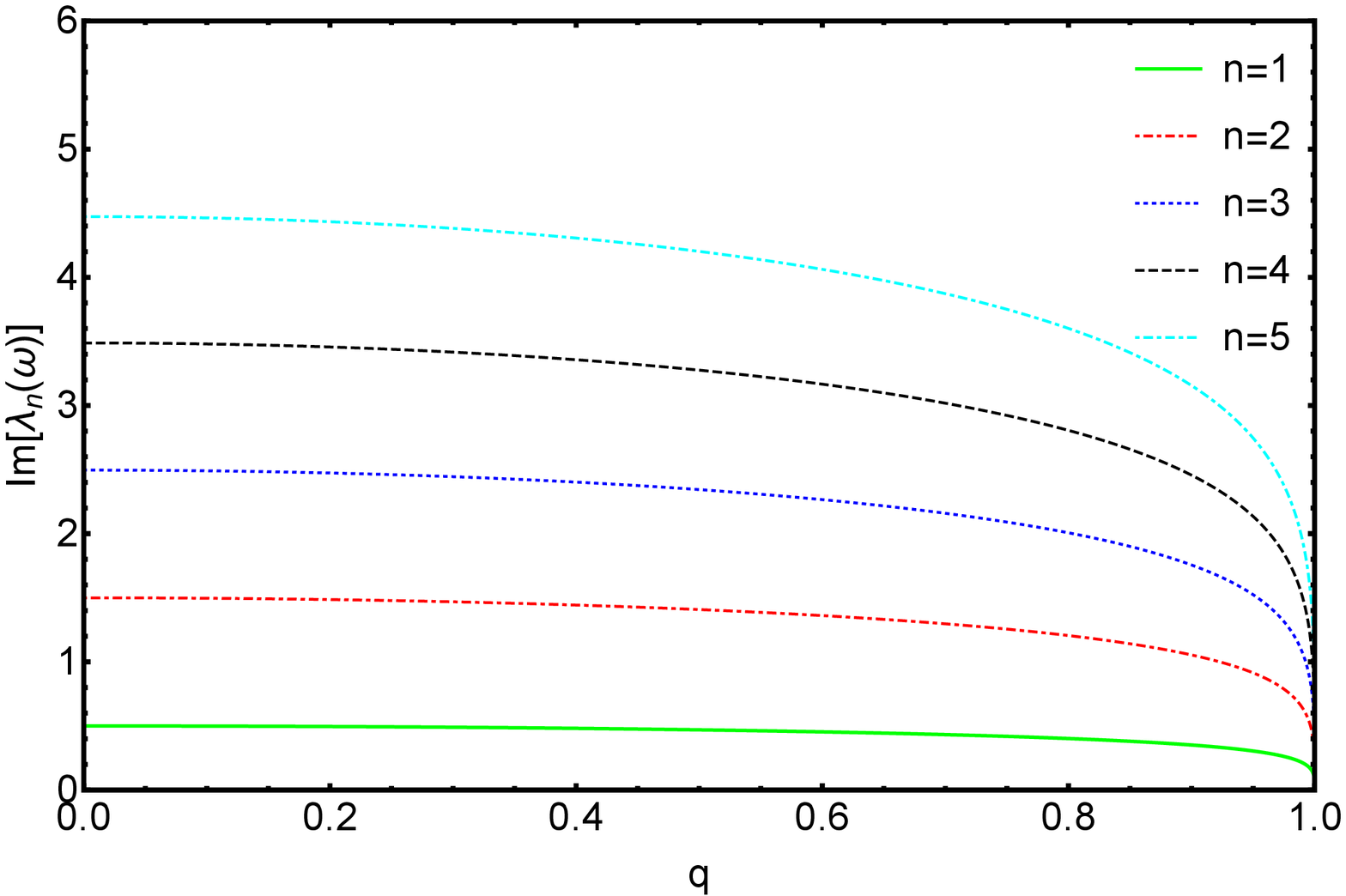}			
	\caption{Real and imaginary parts of the first five Regge poles (with $\omega M=3.0$) as functions of $q$.}
	\label{Fig: regge poles vs q}
\end{figure}

Here, we aim to construct the Regge pole approximation of the scalar scattering cross section of the GMGHS black hole.
To this end, it is necessary to determine the Regge poles $\lambda_{n}(\omega)$ and
the corresponding residues $r_{n}(\omega)$ (see Eq.(\ref{Eq: Regge pole approx})).
We apply the continued fraction method to determine the Regge poles.
By definition, Regge poles are zeros of $A^{(-)}_{\lambda-1/2}(\omega)$ in the complex $\lambda$ plane, they can be obtained
by numerically solving the following equation
\begin{equation}
	0=\beta_{0}-\frac{\alpha_0\gamma_1}{\beta_1-}\frac{\alpha_1\gamma_2}{\beta_2-}\frac{\alpha_2\gamma_3}{\beta_3-}\cdots,
\end{equation}
or alternatively, by solving the $n$-th inversion of this equation.
Here $\alpha_{n},\;\beta_{n}$ and $\gamma_{n}$ are functions of $(\omega,\lambda)$,
and they are given in Eqs.(11)-(13) of Ref.\cite{PhysRevD.92.064022}.
Figure.\ref{Fig: regge poles vs q} shows the real and imaginary parts of the first five Regge poles as functions of $q$.
We see that both real and imaginary parts of $\lambda_n$ decrease monotonically with the increase of $q$.

\begin{figure*}
	\centering
	\subfigure{\label{res: n1}
		\includegraphics[width=0.32\textwidth,height=0.3\textwidth]{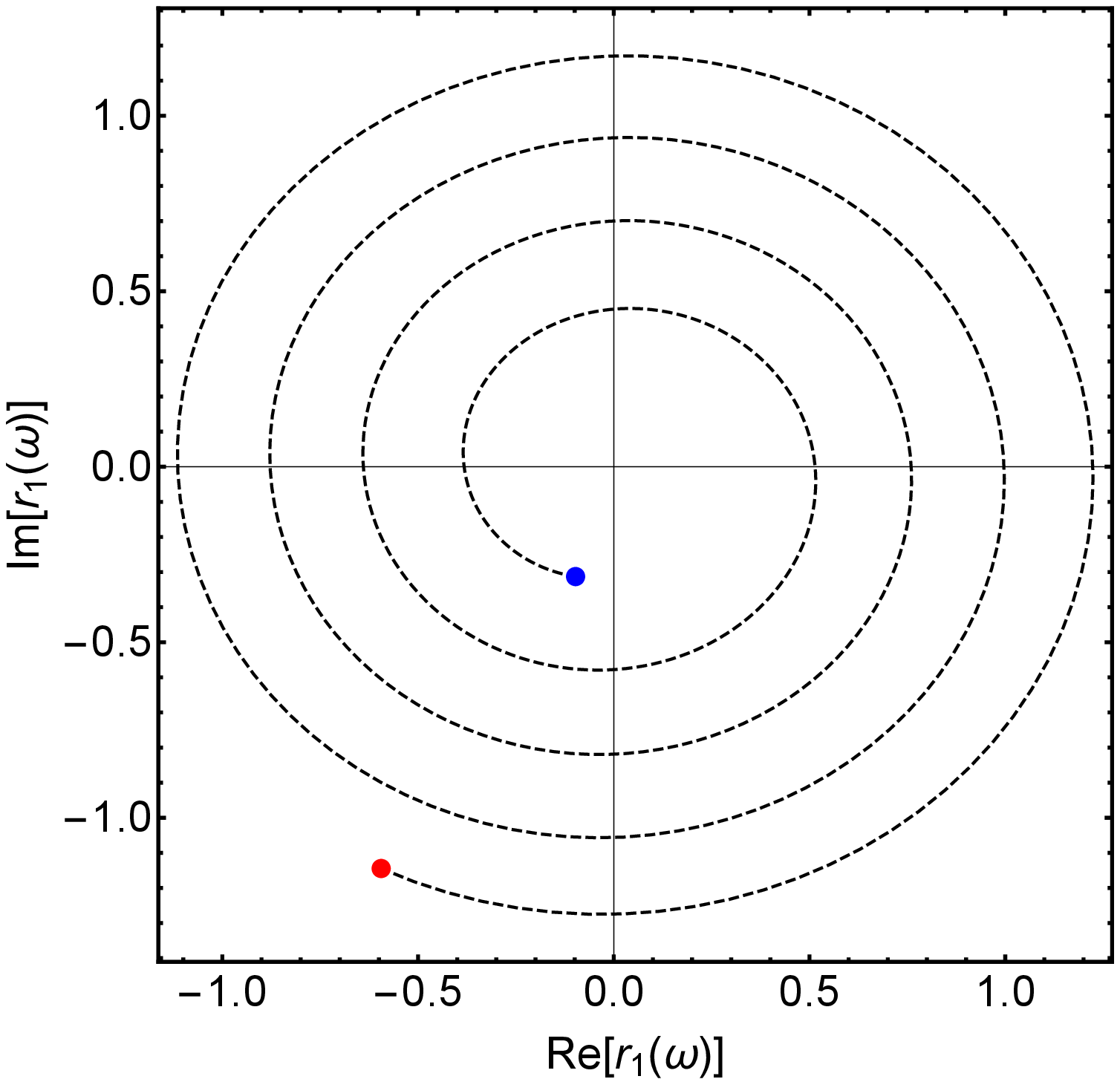}
	}
	\subfigure{\label{res: n2}
		\includegraphics[width=0.32\textwidth,height=0.3\textwidth]{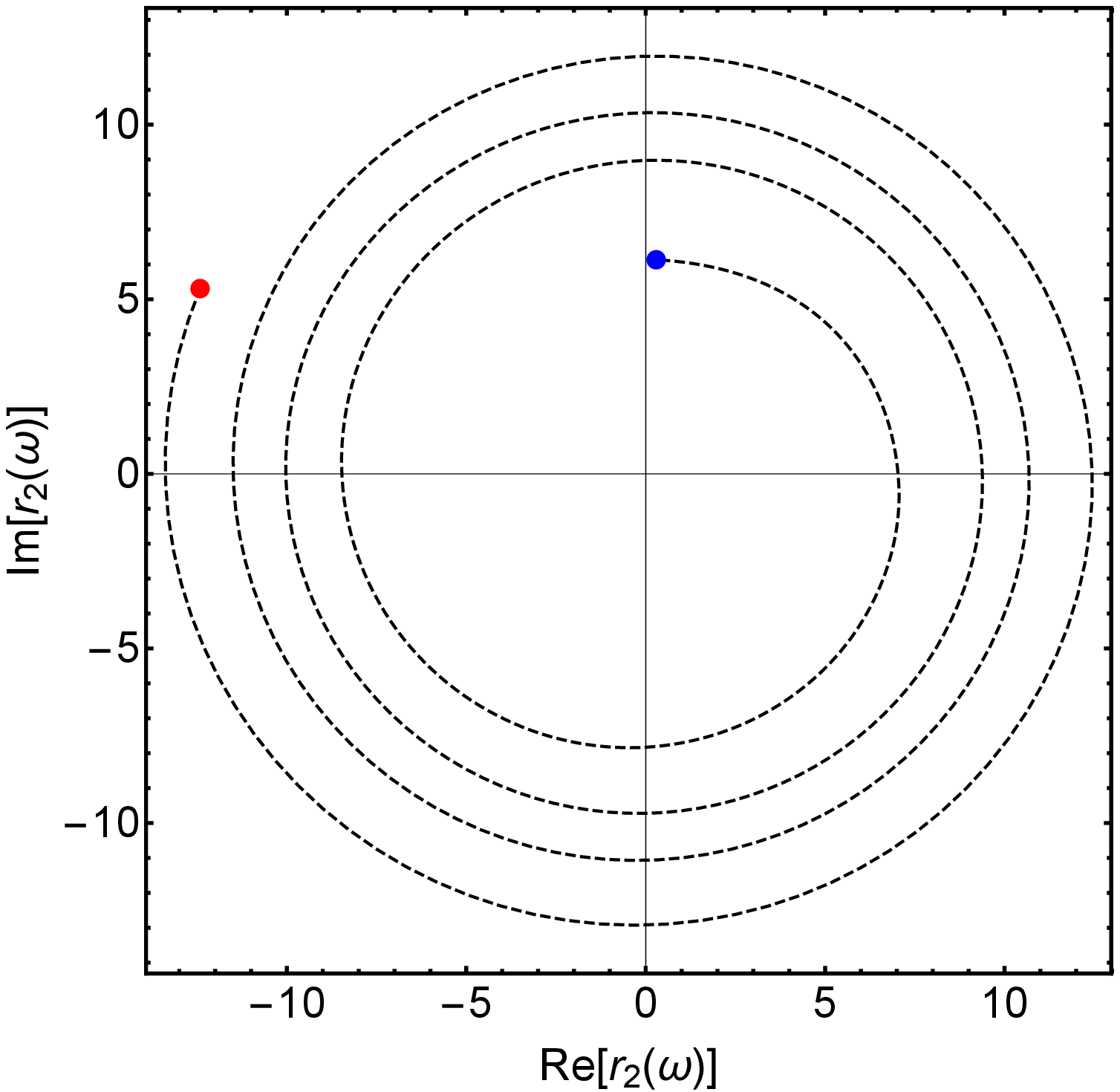}
	}		
	\caption{Path of the first two residues $r_n(\omega)$ as $q$ varies in the range $0\leq q\leq0.999$.
		The red and blue points correspond to $q=0$ and $q=0.999$, respectively.
		We set the frequency $\omega M=3.0$.}
	\label{Fig: res}
\end{figure*}

To get the residues, we apply a method that has been used in Ref.\cite{PhysRevD.88.044018} to compute the quasinormal mode excitation factors.
This method is based on the formalism developed by Mano, Suzuki, and Takasugi (MST) \cite{Mano:1996vt}, see also \cite{Mano:1996gn,Mano:1996mf,Sasaki:2003xr}.
First we compute $A^{(+)}_{\lambda-1/2}(\omega)$ and $A^{(-)}_{\lambda-1/2}(\omega)$ at $\lambda=\lambda_{n}(\omega)$.
Then, we consider $\lambda=\lambda_{n}(\omega)+\delta$ and compute $A^{(-)}_{\lambda-1/2}(\omega)$ at the new value of $\lambda$.
Finally, the derivative of $A^{(-)}_{\lambda-1/2}(\omega)$ with respect to $\lambda$ is given by
\begin{equation}
	\frac{d}{d\lambda}A^{(-)}_{\lambda-1/2}=\frac{A^{(-)}_{\lambda_n+\delta-1/2}-A^{(-)}_{\lambda_n-1/2}}{\delta}.
\end{equation}
In our calculation we choose $\delta=10^{-7}$, i.e., we differentiate along the real axis;
as a check, we also differentiate along the pure-imaginary axis and find that they are consistent in high precision.
In Appendix. \ref{Appendex: MST}, we present the details of how to solve Eq.(\ref{Eq: the radial eq}) and compute
amplitudes $A^{(+)}_{\lambda-1/2}(\omega)$ and $A^{(-)}_{\lambda-1/2}(\omega)$ via the MST method.

In Fig. \ref{Fig: res}, we present the path followed by the real and imaginary parts of the first two residues $r_n(\omega)$, as $q$ varies in the range $0\leq q\leq 0.999$.
Once we have obtained $\lambda_n(\omega)$ and $r_n(\omega)$, we can build up the Regge pole approximation of the scattering amplitude from Eq.(\ref{Eq: Regge pole approx}).

\section{Results}\label{Sec: results}
In this section, we present a selection of the numerical results for the differential scattering cross sections of the GMGHS black hole.

\begin{figure}
	\centering
	\includegraphics[width=0.42\textwidth,height=0.22\textwidth]{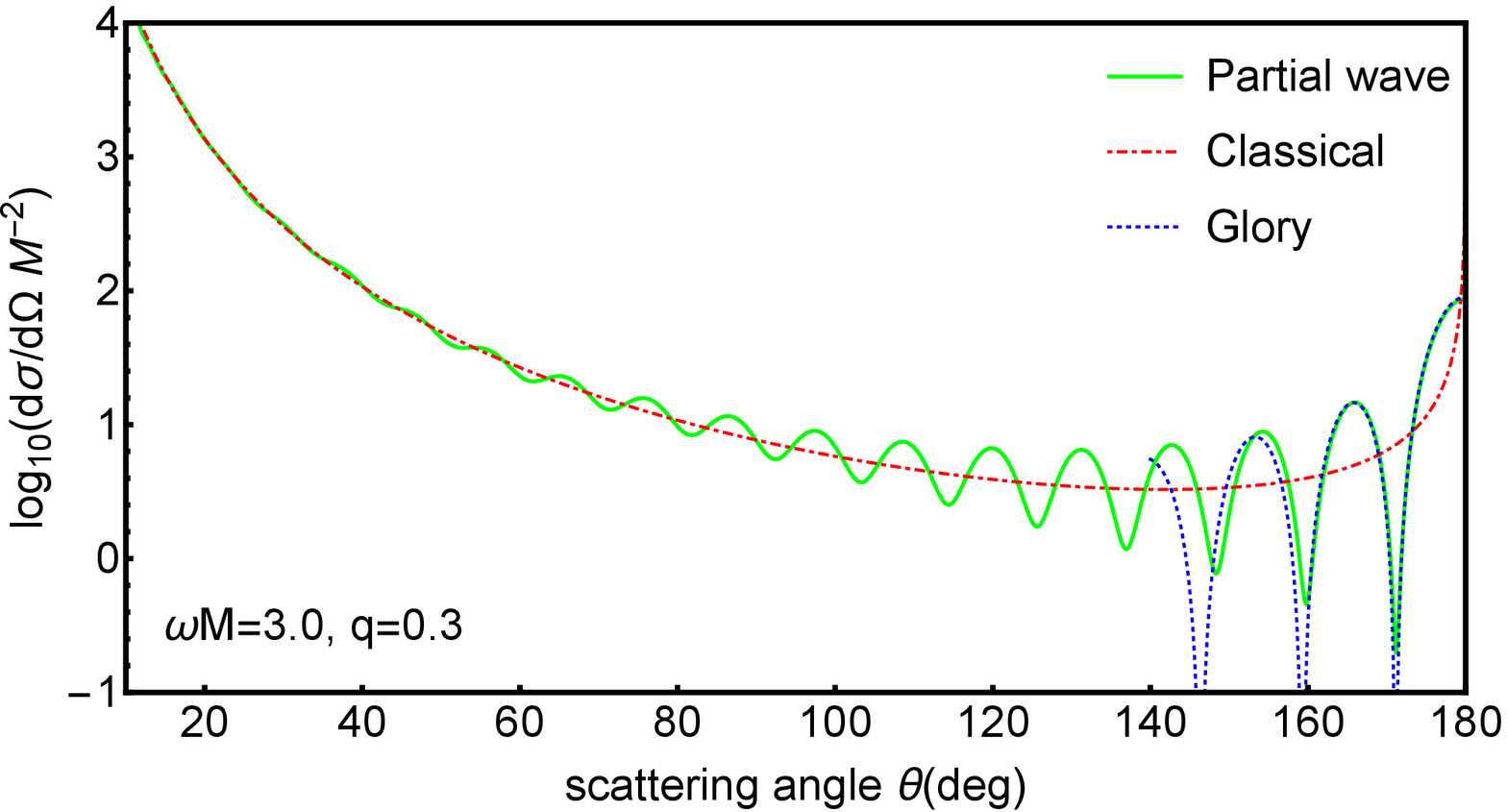}	
	\includegraphics[width=0.42\textwidth,height=0.22\textwidth]{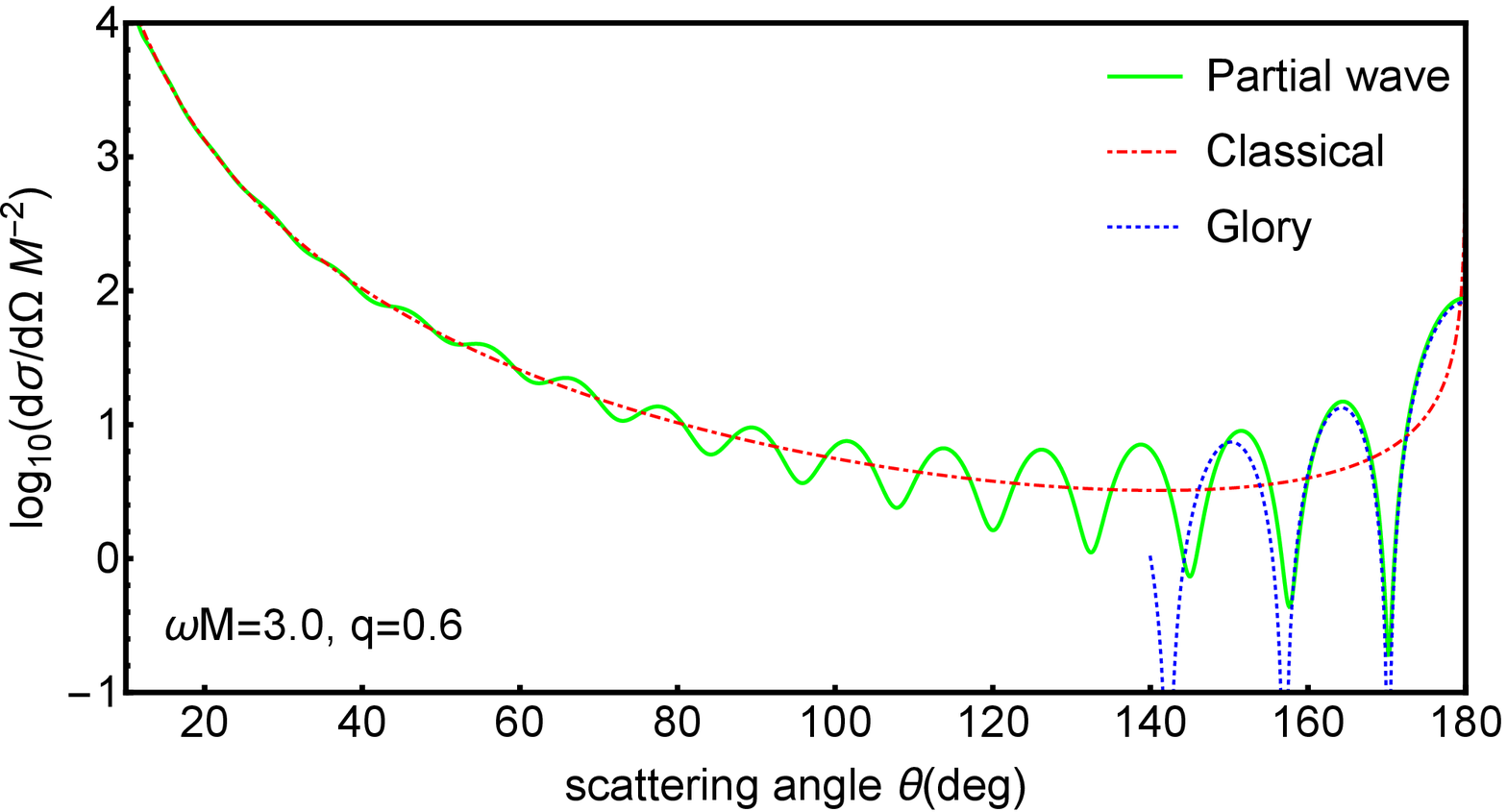}	
	\includegraphics[width=0.42\textwidth,height=0.22\textwidth]{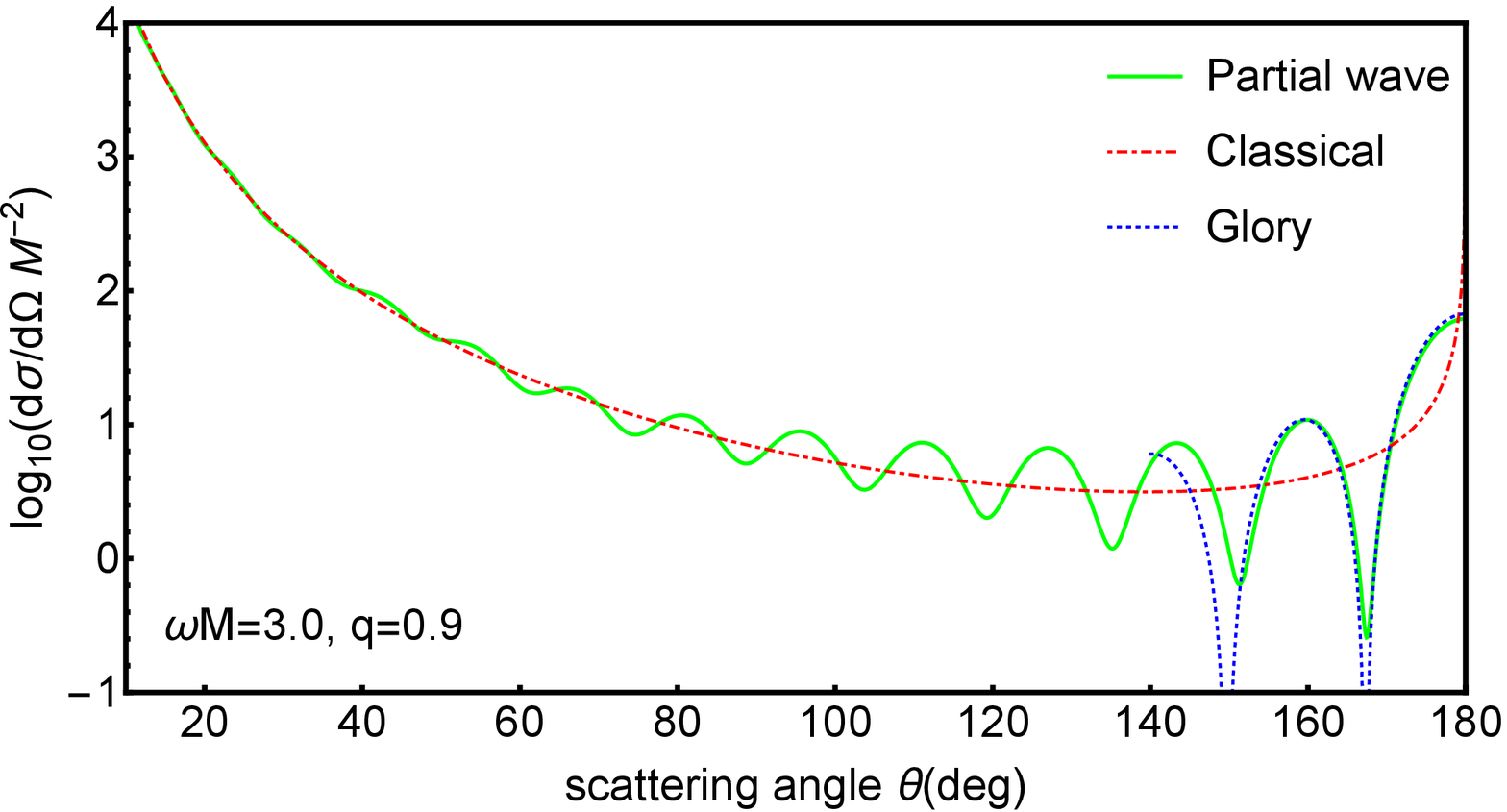}		
	\caption{Comparison of scattering cross sections with the geodesic and glory approximations for $q=0.3$ (top), $0.6$ (middle), and $0.9$ (bottom). We set $\omega M=3.0$.}
	\label{Fig: srcs}
\end{figure}

Figure \ref{Fig: srcs} shows comparisons of the numerical scalar scattering cross sections of the GMGHS black holes
with the classical and semi-classical glory approximations (see Eqs.(\ref{Eq: classical src}) and (\ref{Eq: glory})).
We observe that the glory approximation fits well with the numerical results for large angles ($\theta\gtrsim 160^{\circ}$),
while the classical approximation fits well the small angle region ($\theta\lesssim20^{\circ}$).
In the intermediate range, only full wave scattering treatment yields accurate result.  

\begin{figure}
	\centering
	\includegraphics[width=0.45\textwidth,height=0.29\textwidth]{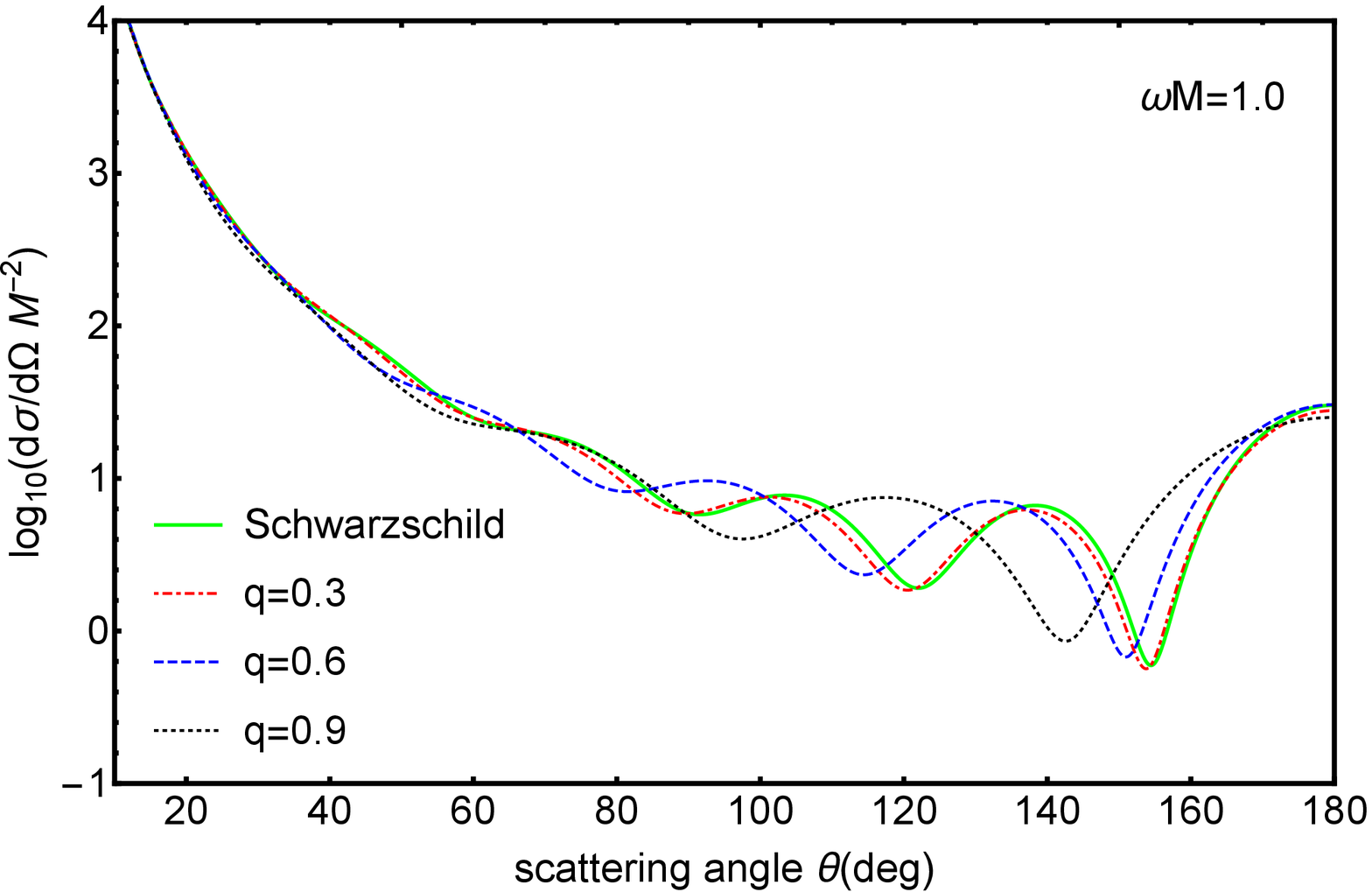}
	\includegraphics[width=0.45\textwidth,height=0.29\textwidth]{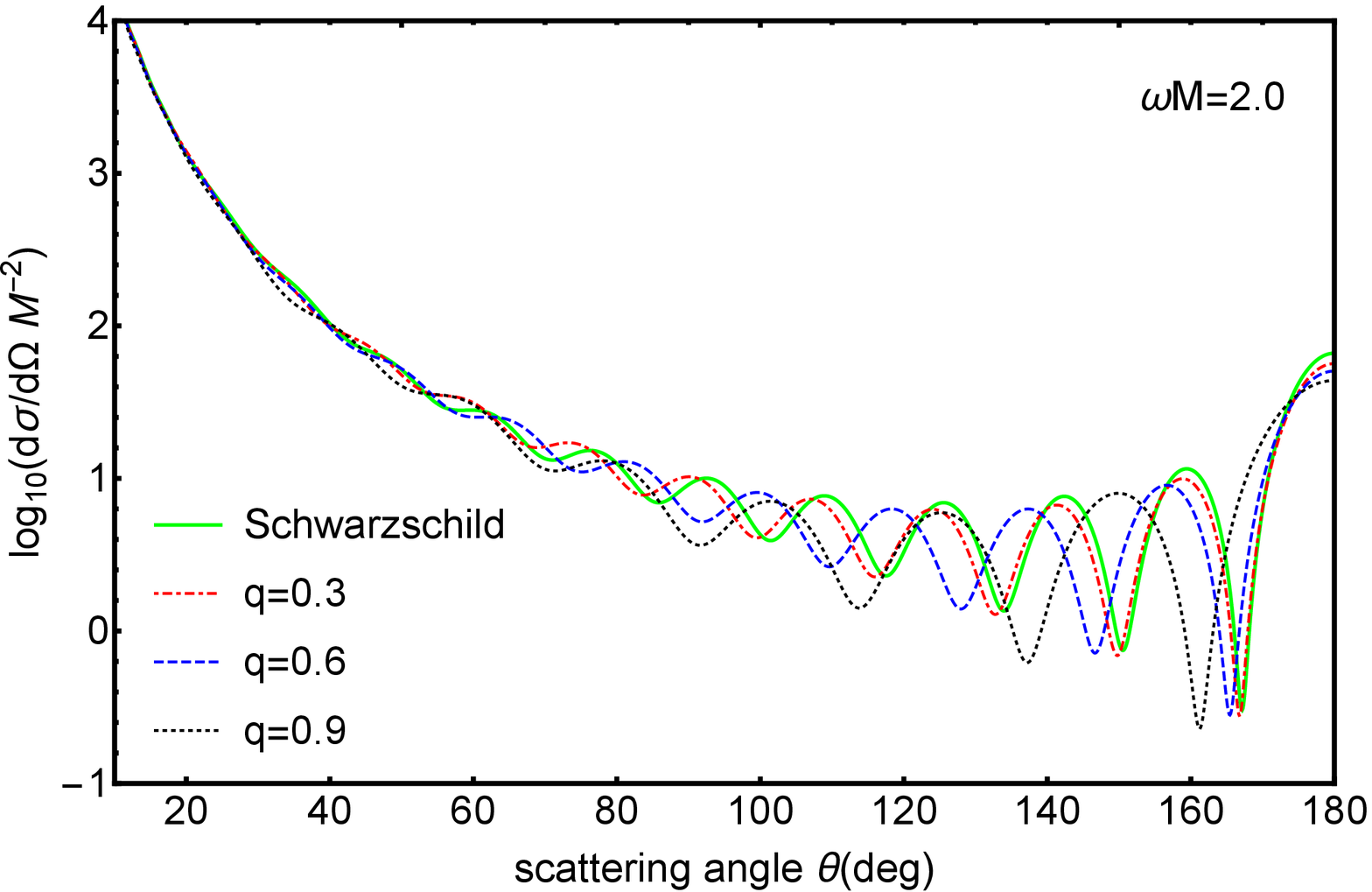}	
	\includegraphics[width=0.45\textwidth,height=0.29\textwidth]{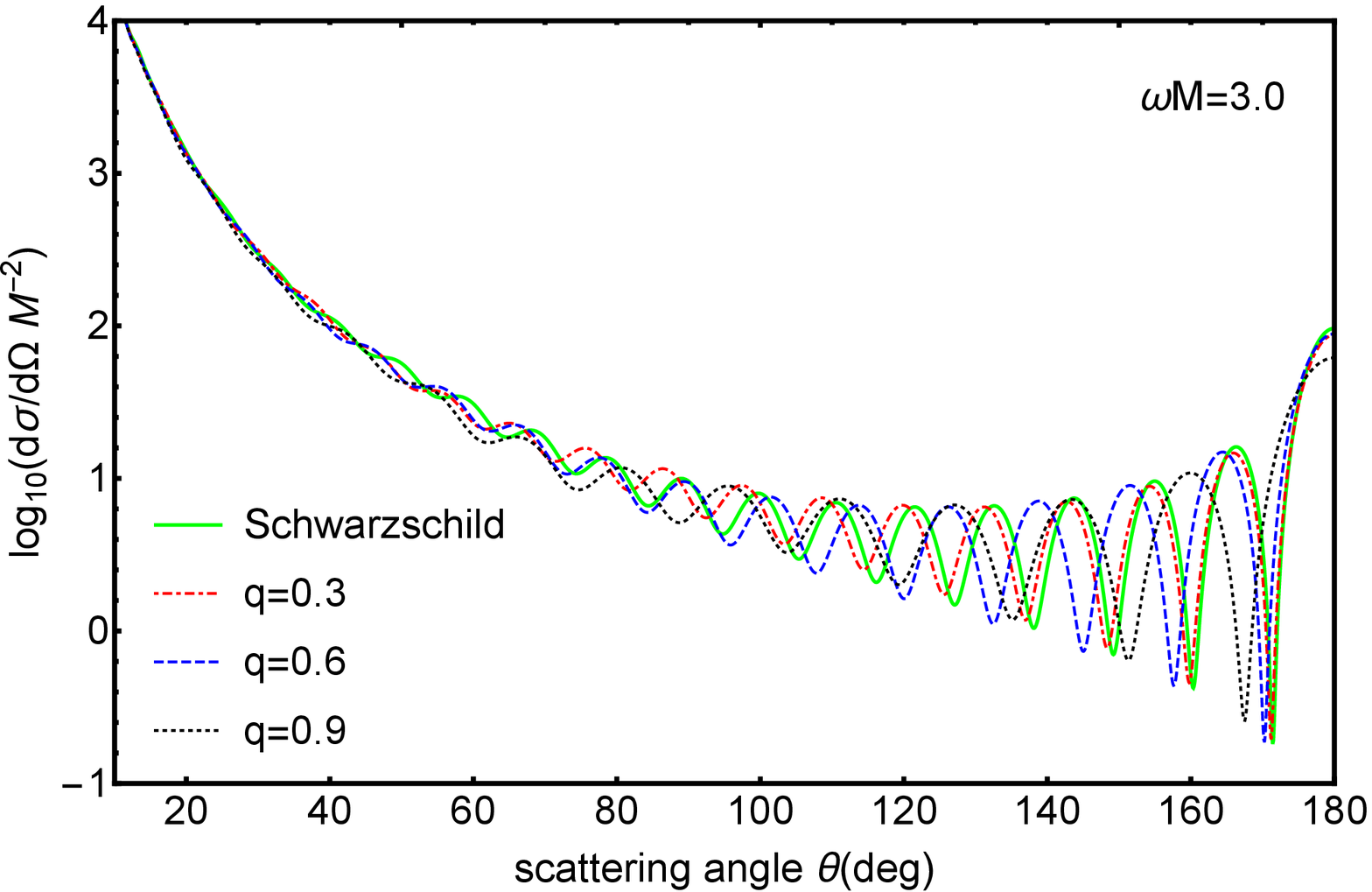}			
	\caption{Comparison of scattering cross sections with $\omega M=1.0$ (top), $2.0$ (middle), and $3.0$ (bottom), for different values of the black hole charge.}
	\label{Fig: src for dif q}
\end{figure}

In Fig. \ref{Fig: src for dif q}, we compare the scattering cross sections for different values of
the black hole charge ($q=0.3,0.6,0.9$) with $\omega M=1.0$ (top), $2.0$ (middle), and $3.0$ (bottom).
For comparison, we also plot the Schwarzschild ($q=0$) case in each subplots of this figure.
From the plots in Fig. \ref{Fig: src for dif q}, we find that (i) the effect of $q$ on the scattering cross section
at small angles is negligible, and (ii) the interference fringe width increases with the increase of the black hole charge $q$
for given values of $\omega M$,
or with the decrease of $\omega M$ for given values of $q$.
These results can be understood as follows:
(i) From Eq.(\ref{Eq: classical src small th}), we see that for small angles,
the scattering cross section is dominated by the term $16M^2/\theta^4$,
the black hole charge $q$ only affects the cross section in the subdominant term proportional to $M^2/\theta^3$;
(ii) The glory approximation given in Eq.(\ref{Eq: glory}) implies that for large angles ($\theta\approx180^{\circ}$),
the interference fringe width is proportional to $1/(b_g\omega)$. In addition, from Fig.\ref{Fig: glory param} we see that
$b_g$ decreases monotonically with the increase of $q$.

\begin{figure}
	\centering
	\includegraphics[width=0.45\textwidth,height=0.29\textwidth]{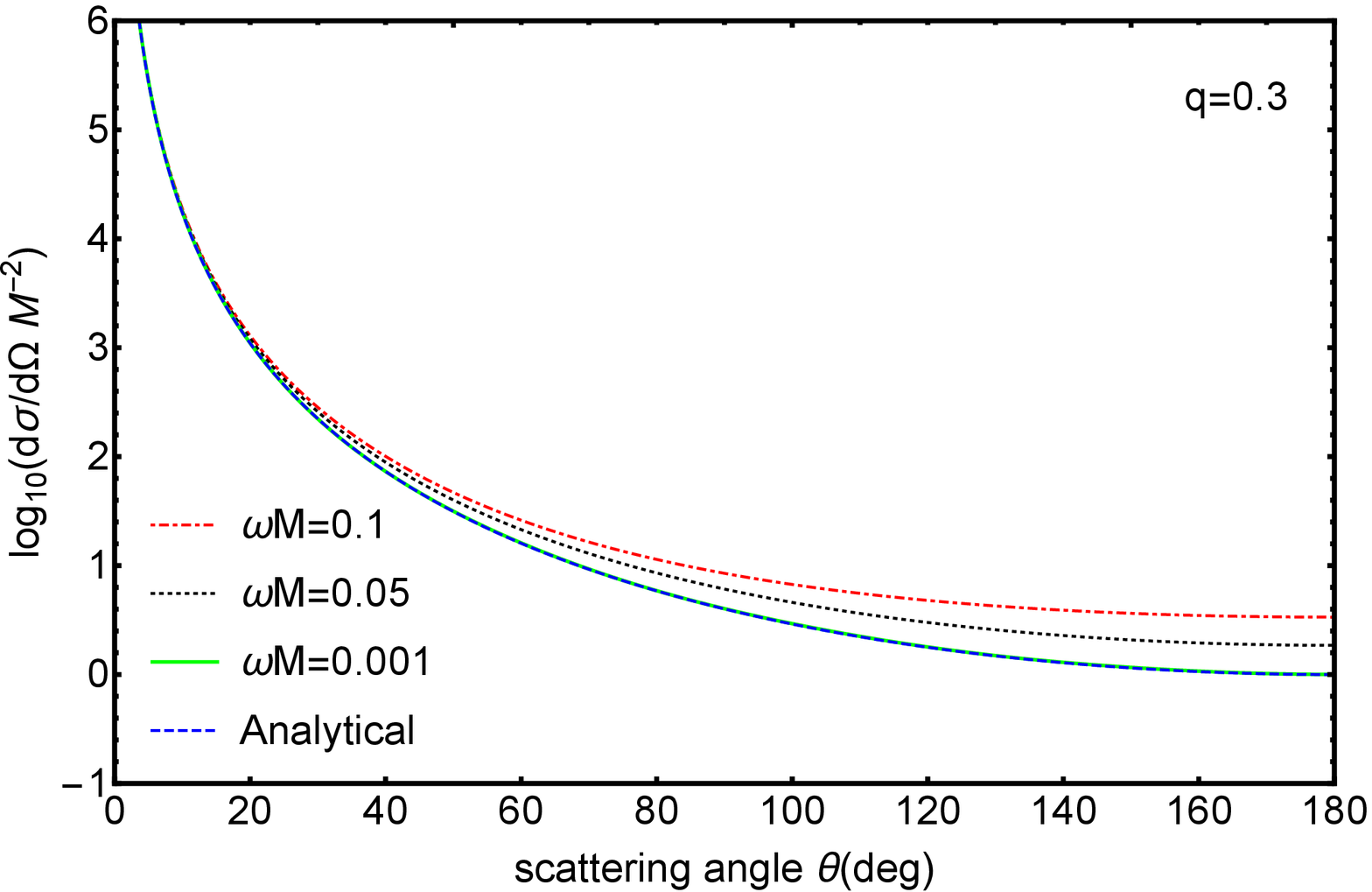}
	\includegraphics[width=0.45\textwidth,height=0.29\textwidth]{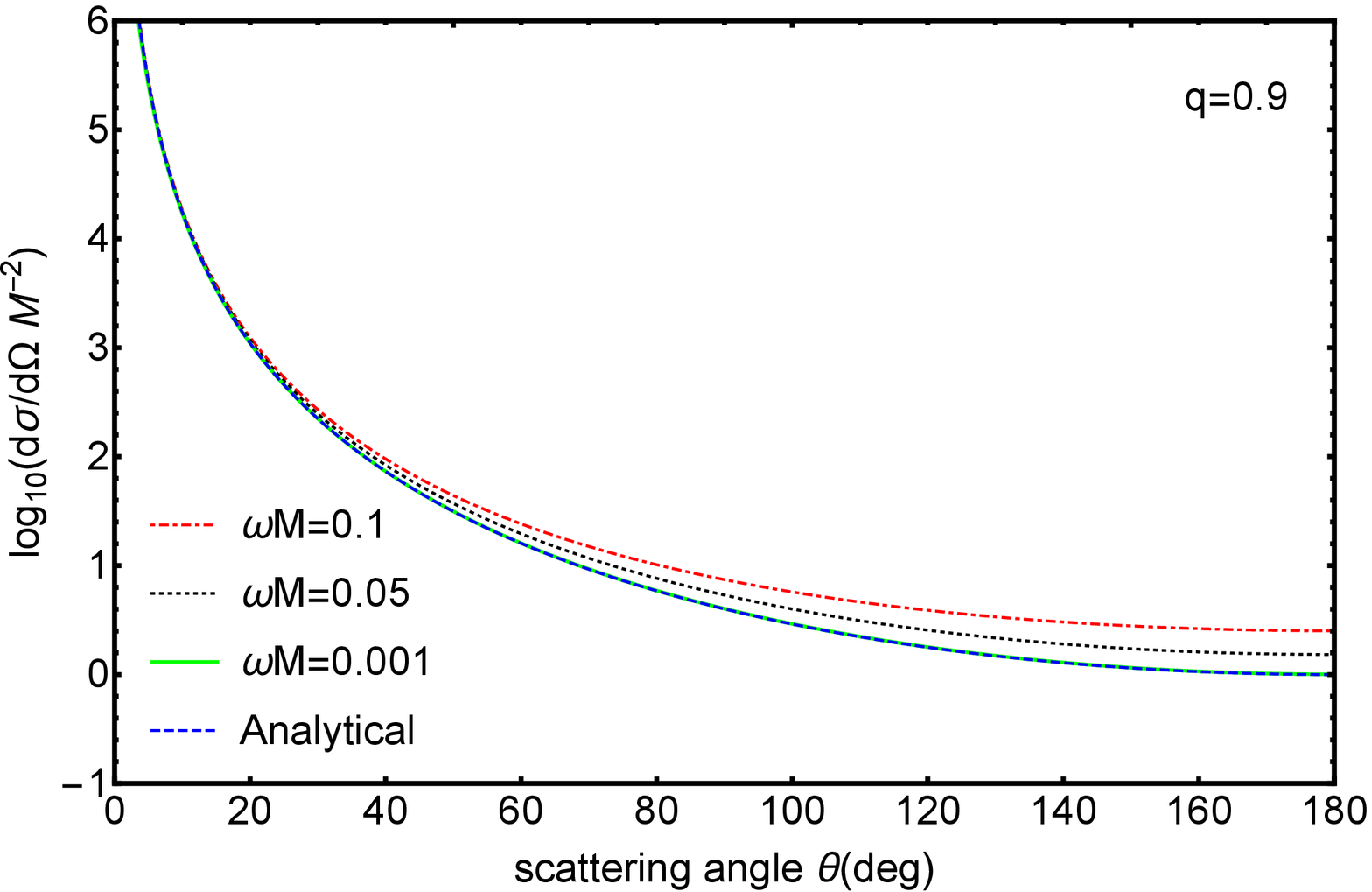}			
	\caption{The low frequency behavior of the scattering cross section for $q=0.3$ (top) and $0.9$ (bottom). 
		     The solid green lines are given by the analytic formula (\ref{Eq: low frequency limit}).
	         This figure shows that in the low frequency limit $\omega M\rightarrow0$, the scattering cross section of the GMGHS black hole is independent on $q$.}
	\label{Fig: low frequency}
\end{figure}

Figure \ref{Fig: low frequency} shows the scalar scattering cross sections of the GMGHS black holes at low frequencies. 
In this case, the scattering cross section does not oscillate along the $\theta$-axis, and the scattering cross section decreases with the frequency.
Furthermore, in the limit $\omega M\rightarrow 0$, the scattering cross section of the GMGHS black hole is given by \cite{PhysRevD.13.775}
\begin{equation}\label{Eq: low frequency limit}
	\lim\limits_{\omega M\rightarrow 0}\left(\frac{1}{M^2}\frac{d\sigma}{d\Omega}\right)=\frac{1}{\sin^4(\theta/2)}.
\end{equation}

\begin{figure}
	\centering
	\includegraphics[width=0.45\textwidth,height=0.32\textwidth]{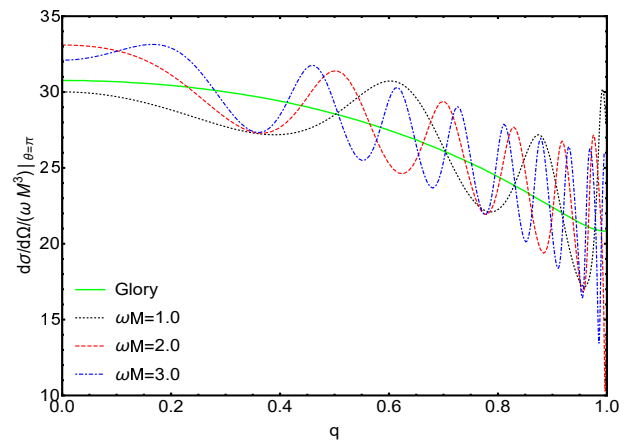}		
	\caption{Amplitudes of the backscattered waves ($\theta=180^{\circ}$) as functions of $q$ for different values of $\omega M$.
	         The solid green line is given by the glory approximation.}
	\label{Fig: bcs}
\end{figure}

Although the glory approximation fits the scattering cross section remarkably well for large angles,
there is a quantitative difference between the amplitudes of the backscattered wave ($\theta=180^{\circ}$) obtained via the partial wave
method and the glory approximation \cite{PhysRevD.79.064022,PhysRevD.92.024012}.
In Fig. \ref{Fig: bcs}, we compare the amplitudes of the backscattered wave
obtained via the glory approximation and the partial wave method. 
We see that the results obtained via partial wave method oscillate around the semi-classical glory result.

\begin{figure}
	\centering
	\includegraphics[width=0.45\textwidth,height=0.29\textwidth]{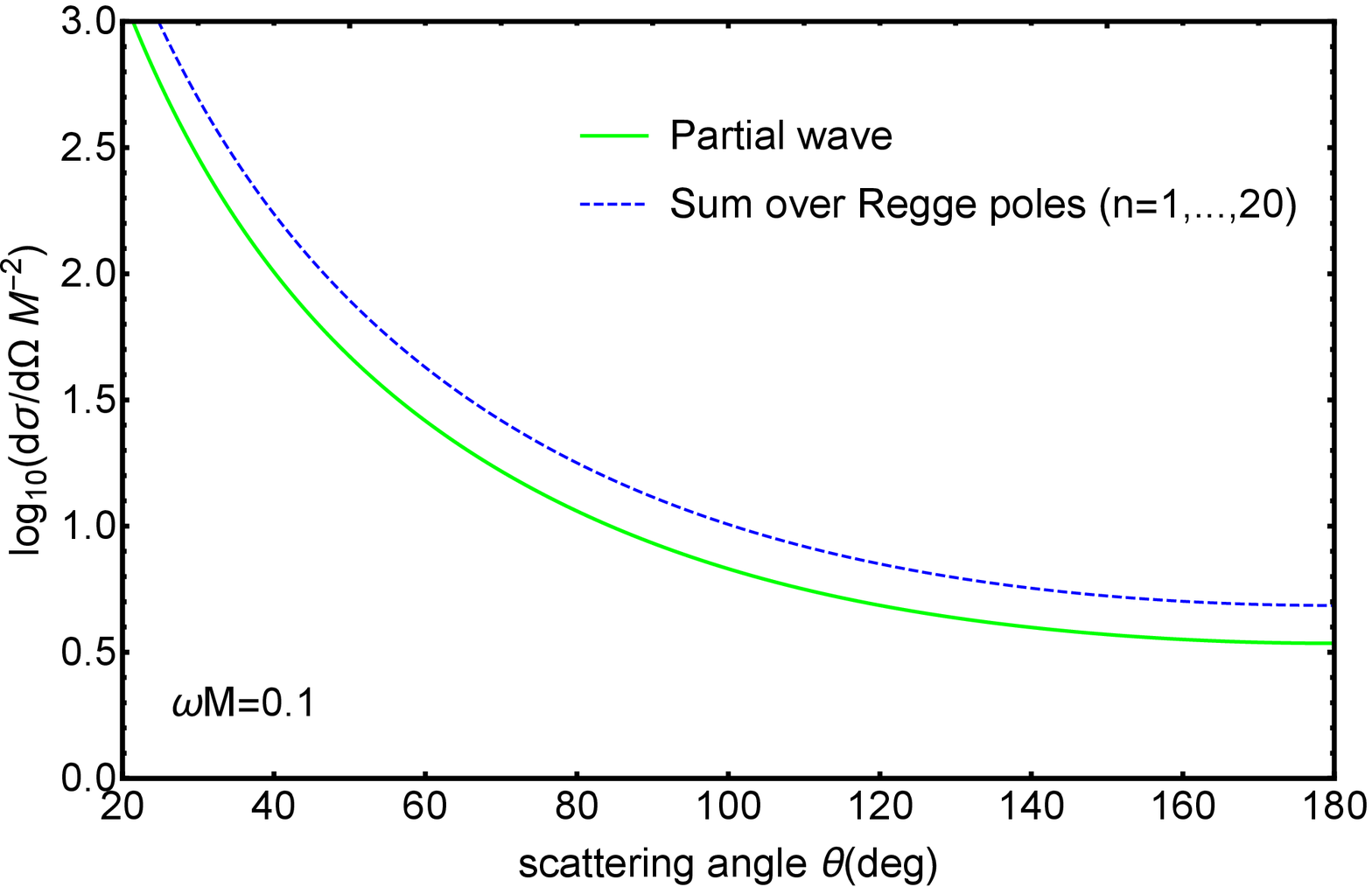}
	\includegraphics[width=0.45\textwidth,height=0.29\textwidth]{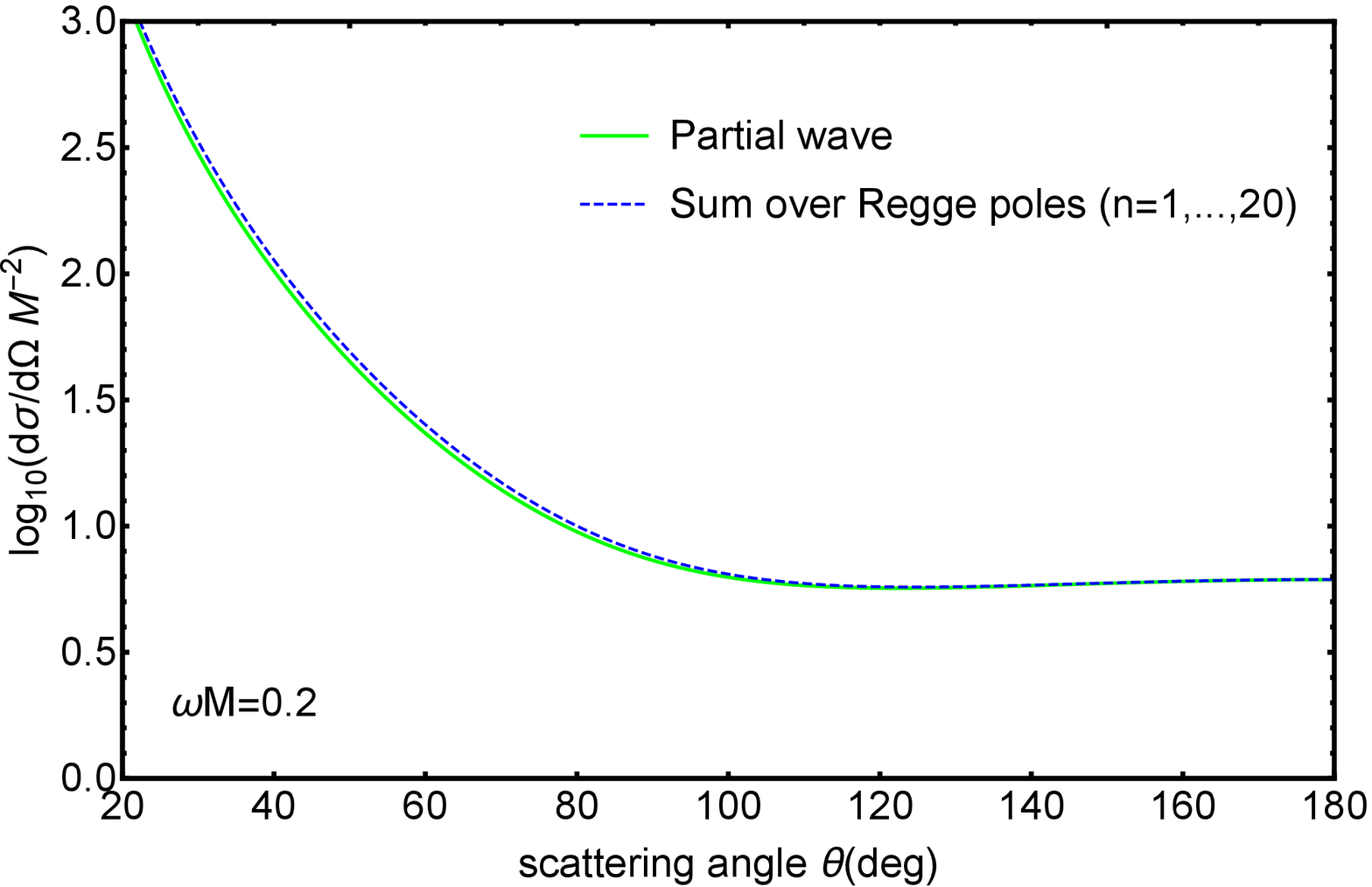}	
	\includegraphics[width=0.45\textwidth,height=0.29\textwidth]{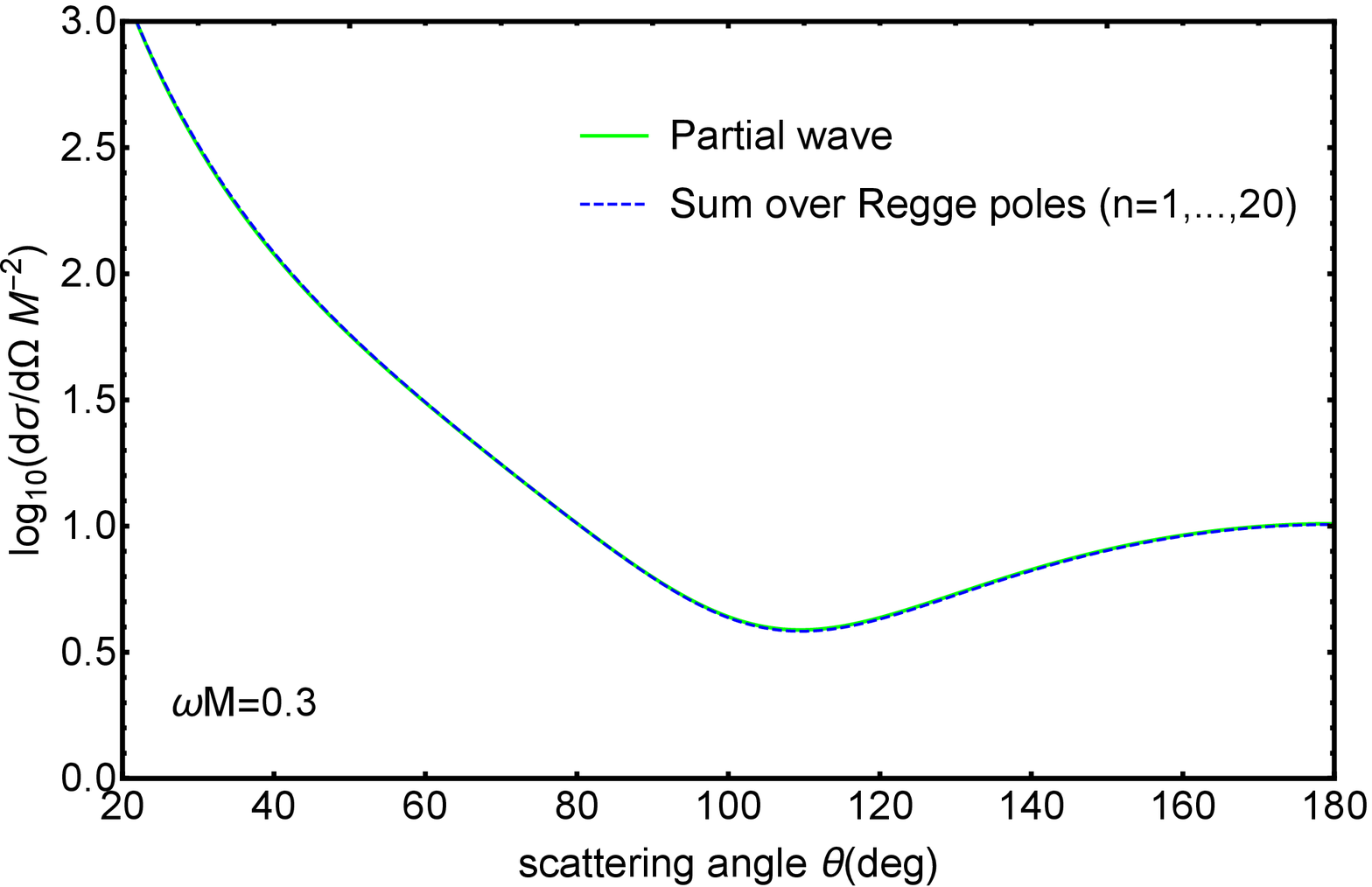}			
	\caption{Comparison of the scattering cross sections obtained via the partial wave method and the Regge pole approximations for $\omega M=0.1$ (top), $0.2$ (middle), and $0.3$ (bottom). We set $q=0$.}
	\label{Fig: src regge without background}
\end{figure}

Let us now focus on the Regge pole approximation of the scattering cross section. 
In Fig. \ref{Fig: src regge without background}, we compare the partial wave results of the scattering cross section with its Regge pole approximation
at low frequencies, i.e. $\omega M\leq0.3$. This plot shows clearly that the difference between the partial wave result and the Regge pole approximation decreases with the increase of $\omega M$.
We evaluate the relative error of the Regge pole approximation in the range $20^\circ\leq\theta\leq180^\circ$, and find that for $\omega M=0.1$ the maximum relative error is about $42.6\%$, 
whereas for $\omega M=0.3$ is only about $2.1\%$. This is consistent with the results in \cite{Folacci:2019cmc} that the contribution from the background integral for high frequencies is negligible.

\begin{figure}
	\centering
	\includegraphics[width=0.45\textwidth,height=0.29\textwidth]{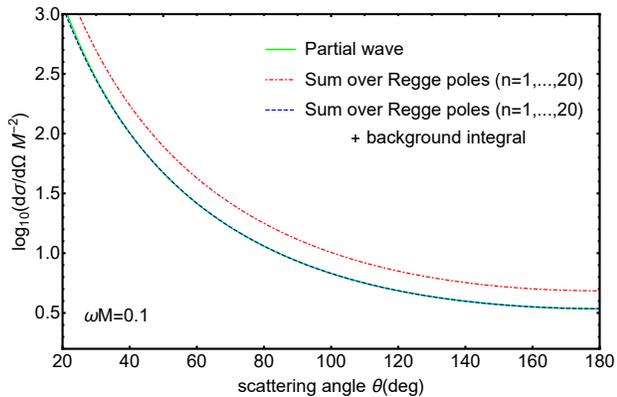}		
	\caption{Comparison of the partial wave result of the scattering cross section with its Regge pole approximation with and without the background integral. The parameters are the same as those in Fig. \ref{Fig: src regge without background} (with $\omega M=0.1$).}
	\label{Fig: src regge with background}
\end{figure}

In Fig.\ref{Fig: src regge with background}, we plot the scattering cross section obtained from the partial wave method and the Regge pole approximation with and without the background integral\footnote{We apply the series reduction technique introduced in \cite{Folacci:2019cmc} to improve the convergence of the background integral.}.
We can see that taking into account the contribution from the background integral improves the Regge pole approximation drastically at low frequencies.

\begin{figure}
	\centering
	\includegraphics[width=0.45\textwidth,height=0.3\textwidth]{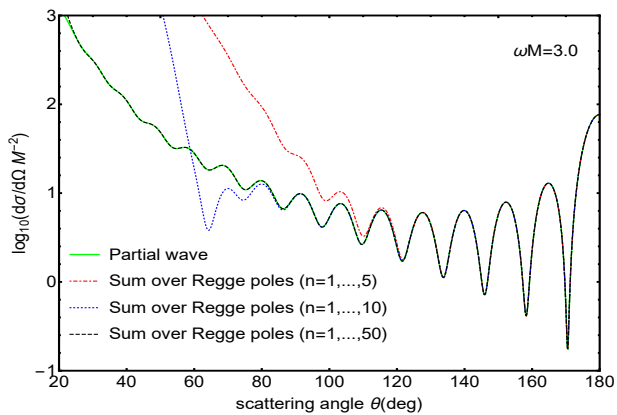}		
	\caption{Comparison of the scattering cross sections obtained via the partial wave method and the Regge pole approximations.
	         The black hole charge is $q=0.3$.}
	\label{Fig: src regge}
\end{figure}

Figure \ref{Fig: src regge} compares the scattering cross sections obtained by the partial wave method and the Regge pole approximation for $q=0.3$.
The frequency is $\omega M=3.0$.
We see that the Regge pole approximation (with only $5$ Regge poles) is very close to the result obtained via partial wave method
for a wide range of angles ($\theta\gtrsim120^{\circ}$), and the Regge pole approximation can be improved by taking into account more Regge poles in the sum (\ref{Eq: Regge pole approx});
By summing over $50$ Regge poles, we get the result that is indistinguishable from the partial wave result for $\theta\gtrsim20^{\circ}$.

\begin{figure}
	\centering
	\includegraphics[width=0.45\textwidth,height=0.3\textwidth]{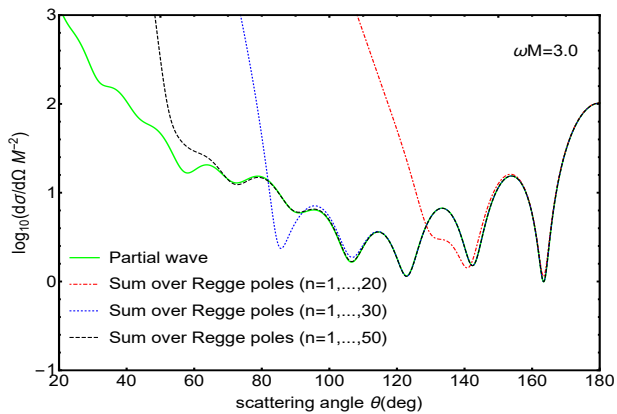}		
	\caption{Comparison of the scattering cross sections obtained via the partial wave method and the Regge pole approximations.
			 The black hole charge is $q=0.999$.}
	\label{Fig: src regge_q999}
\end{figure}

In Fig. \ref{Fig: src regge_q999}, we compare the Regge pole approximation with the partial wave result of
the scattering cross section of a near extremal GMGHS black hole ($q=0.999$).
Again, the Regge pole approximations fit the partial wave result well for large angles.
However, we find that with the increase of $q$, more Regge poles should be taken into account in order to obtain results at the same precision.
As is shown Fig. \ref{Fig: src regge_q999}, summing over $50$ Regge poles yields the result in good agreement with the partial wave method for $\theta\gtrsim80^{\circ}$.

\begin{figure}
	\centering
	\includegraphics[width=0.45\textwidth,height=0.32\textwidth]{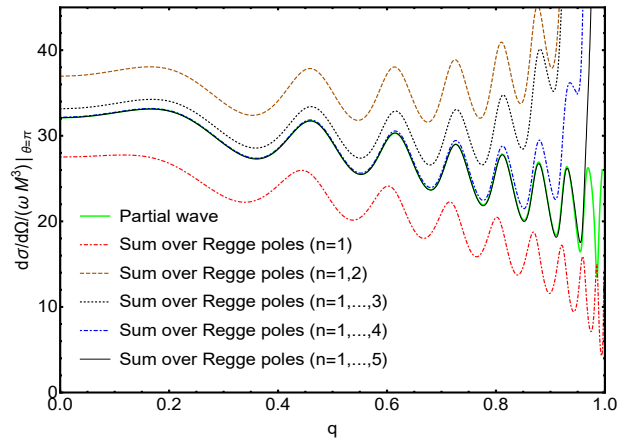}		
	\caption{Comparison of the amplitudes of the backscattered waves obtained by the partial wave method and the sum over Regge poles.}
	\label{Fig: intensity}
\end{figure}

Figure \ref{Fig: intensity} compares the amplitudes of the backscattered wave obtained
via the partial wave method and Regge pole approximation for $\omega M=3.0$.
The plot shows that summing over a few Regge poles captures most of the features of the the backscattered wave.
With only $4$ Regge poles, the agreement is excellent for $q\lesssim0.7$.
When $q$ increases, more Regge poles should be included to describe the backscattered waves.
This is consistent with the result in Fig. \ref{Fig: src regge_q999}.

\section{Conclusion}\label{Sec: conclusion}

Scattering from black holes is a fundamental and important problem in astrophysics and theoretical physics. Because of its ultrastrong attractive interaction, black holes stimulate several novel phenomena  
which are never seen in scattering experiment in lab, for example black hole shadow, glory etc. Some objects, for example the gravitational wave, neutrino and the expected dark matter particles, only have 
extremely weak interaction with ordinary matters. Thus it is difficult to observe their scattering phenomena with ordinary matters. On the contrary, they strongly interact with black holes and yield conspicuous phenomena \cite{Lin:2019qyx,Hongsheng:2018ibg}. To study the scattering from black holes is very helpful to explore the elusive objects, including gravitational wave, neutrino and dark matters.

In this paper we discuss the scattering of a massless scalar field by the GMGHS black hole.
Using the partial wave method, we numerically computed the scalar scattering cross sections,
and compared the numerical results with the geodesic and glory approximations.
We summarize the main results as follows:

First, the scattering cross section is well approximated by the classical results obtained via the geodesic analysis for small angles,
and the effects of the black hole charge is negligible in this case (see Eq.(\ref{Eq: classical src small th})).

Second, for large scattering angles, the black hole charge has a significant effect on the scattering cross section, which is well described by the glory approximation.
We showed that the interference fringe width increases with the increase of the black hole charge, which is predicted by the glory approximation.
However, the amplitude of the backscattered wave is not equal to the glory amplitude.
We showed that the numerical result oscillates around the glory amplitude when $q$ increases, see Fig. \ref{Fig: bcs}.

We also construct the scattering cross sections of the GMGHS black holes by sum over Regge poles.
We find elegant agreement of the Regge pole analysis and the partial wave results, especially for intermediate and large angles.
Finally, we show that when $q$ increases, it is necessary to include more Regge poles to derive the scattering cross section.
\begin{acknowledgments}
	We thank the anonymous reviewer for their helpful comments and suggestions. Y. Huang wishes to thank Yuan-Xing Gao for helpful discussions. This work is supported by the National Natural Science Foundation of China (NSFC) under grant Nos.11805166 and 11575083, as well as Shandong Province Natural Science Foundation under grant No. ZR201709220395.
\end{acknowledgments}

\appendix
\begin{widetext}
\section{Computing $A^{(+)}_{\lambda-1/2}(\omega)$ and $A^{(-)}_{\lambda-1/2}(\omega)$}\label{Appendex: MST}

In this appendix, we present the details of how to solve the radial equation (\ref{Eq: the radial eq}) and compute
amplitudes $A^{(+)}_{\lambda-1/2}(\omega)$ and $A^{(-)}_{\lambda-1/2}(\omega)$ via the MST method \cite{Mano:1996mf,Mano:1996vt,Mano:1996gn}.

\subsection{Near horizon solution in series of hypergeometric functions}\label{Subsec: near horizon}
It is convenient to express the radial equation in terms of the dimensionless variables
\begin{equation}\label{Eq: x, kappa and eps}
x=-\frac{r-r_+}{\kappa r_+};\;\;\;\kappa=1-\frac{Q^2}{2M^2};\;\;\;\epsilon=2M\omega,
\end{equation}
where $r_+=2M$. Then, the radial equation (\ref{Eq: the radial eq}) becomes
\begin{equation}\label{Eq: the radial eq to x}
x(1-x)\frac{d^2R}{dx^2}+(1-2x)\frac{dR}{dx}+U R=0,
\end{equation}
where
\begin{equation}
U=\frac{\epsilon^2}{x}+l(l+1)-\left(1+2\kappa\right)\epsilon^2+\left(2+\kappa\right)\kappa\epsilon^2 x-\kappa^2\epsilon^2 x^2.
\end{equation}
For simplicity, we have omitted the subscripts "$\omega l$" of the radial function. The ingoing-wave radial solution has the following form
\begin{equation}\label{Eq: ingoing soln}
R^{\text{in}}(x)=e^{i\kappa\epsilon x}(-x)^{-i\epsilon}p_{\text{in}}(x).
\end{equation}
Substituting Eq.(\ref{Eq: ingoing soln}) into Eq.(\ref{Eq: the radial eq to x}), one finds the differential equation of $p_{\text{in}}(x)$

		\begin{equation}\label{Eq: eq of p(x)}
		x(1-x)p_{\text{in}}''+\left[1-2i\epsilon-(2-2i\epsilon)x\right]p_{\text{in}}'+\left[l(l+1)+i\epsilon(1-i\epsilon)\right]p_{\text{in}}=2i\kappa\epsilon\left[xp_{\text{in}}-x(1-x)p_{\text{in}}'\right]+\epsilon(\epsilon-i\kappa)p_{\text{in}},
		\end{equation}
	where a prime denotes $d/dx$. The left-hand side of Eq.(\ref{Eq: eq of p(x)}) is in the form of a hypergeometric equation. In the limit $\epsilon\rightarrow0$, a solution of Eq.(\ref{Eq: eq of p(x)}) that is finite at $x=0$ is given by
	\begin{equation}
	p_{\text{in}}(\epsilon\rightarrow 0)=F(-l,l+1;1;x).
	\end{equation}
	For a general value of $\epsilon$, the solution of Eq.(\ref{Eq: eq of p(x)}) may be expanded in a series of hypergeometric functions with $\epsilon$ being a kind of expansion parameter.
	The essential point of the MST formalism is to introduce the so-called renormalized angular momentum $\nu$. By adding the term $\left[\nu(\nu+1)-l(l+1)\right]p_{\text{in}}$ to both sides of Eq.(\ref{Eq: eq of p(x)}), one obtains
		\begin{equation}\label{Eq: eq of p(x) nu}
		\begin{aligned}
		x(1-x)p_{\text{in}}''+\left[1-2i\epsilon-(2-2i\epsilon)x\right]p_{\text{in}}'+&\left[\nu(\nu+1)+i\epsilon(1-i\epsilon)\right]p_{\text{in}}\\&=2i\kappa\epsilon\left[xp_{\text{in}}-x(1-x)p_{\text{in}}'\right]+\left[\nu(\nu+1)-l(l+1)+\epsilon(\epsilon-i\kappa)\right]p_{\text{in}}.
		\end{aligned}
		\end{equation}
	Clearly, for arbitrary value of $\nu$, the above equation is equivalent to Eq.(\ref{Eq: eq of p(x)}).
	The trick is to consider the right-hand side of Eq.(\ref{Eq: eq of p(x) nu}) as a perturbation,
	and look for a formal solution specified by the index $\nu$ in a series expansion form
	\begin{equation}\label{Eq: series Hyper}
	p^{\nu}_{\text{in}}(x)=\sum_{n=-\infty}^{\infty}a_{n}^{\;\nu}p_{n+\nu}(x),
	\end{equation}
	where
	\begin{equation}
	p_{n+\nu}(x)=F(n+\nu+1-i\epsilon,-n-\nu-i\epsilon;1-2i\epsilon;x).
	\end{equation}
	Then, we look for the eigenvalue of $\nu$ such that the series expansion is convergent.
	We also require that in the limit $\epsilon\rightarrow0$, the generalized angular momentum tends to $l$.
	This eigenvalue problem could be solved via the continued fraction method.
	It is easy to show that the hypergeometric functions $p_{n+\nu}(x)$ satisfy the recurrence relations
		\begin{equation}
		\begin{aligned}
		xp_{n+\nu}=&-\frac{(\alpha+n)(n+\gamma-\beta)}{(\alpha-\beta+2n)(\alpha-\beta+2n+1)}p_{n+\nu+1}+\frac{2(\alpha+n)(n-\beta)+(\alpha+\beta-1)\gamma}{(\alpha-\beta+2n-1)(\alpha-\beta+2n+1)}p_{n+\nu}
		\\&-\frac{(n-\beta)(\alpha-\gamma+n)}{(\alpha-\beta+2n-1)(\alpha-\beta+2n)}p_{n+\nu-1},
		\end{aligned}
		\end{equation}
		\begin{equation}
		\begin{aligned}
		x(1-x)p_{n+\nu}'=&\frac{(\alpha+n)(n-\beta)(n+\gamma-\beta)}{(\alpha-\beta+2n)(\alpha-\beta+2n+1)}p_{n+\nu+1}+\frac{(\alpha+n)(n-\beta)(\alpha+\beta-2\gamma+1)}{(\alpha-\beta+2n-1)(\alpha-\beta+2n+1)}p_{n+\nu}
		\\&-\frac{(\alpha+n)(n-\beta)(\alpha-\gamma+n)}{(\alpha-\beta+2n-1)(\alpha-\beta+2n)}p_{n+\nu-1}.
		\end{aligned}
		\end{equation}
	where $\alpha=\nu+1-i\epsilon$, $\beta=-\nu-i\epsilon$ and $\gamma=1-2i\epsilon$.
	Substituting Eq.(\ref{Eq: series Hyper}) into Eq.(\ref{Eq: eq of p(x) nu}) and using the above recurrence relations,
	we obtain a three-term recurrence relation among the expansion coefficients $a_{n}^{\;\nu}$
	\begin{equation}\label{Eq: recurrence an}
	\alpha_{n}^{\;\nu}a_{n+1}^{\;\nu}+\beta_{n}^{\;\nu}a_{n}^{\;\nu}+\gamma_{n}^{\;\nu}a_{n-1}^{\;\nu}=0,
	\end{equation}
	where
		\begin{subequations}\label{Eq: alpha_n}
			\begin{empheq}{align}
			&\alpha_{n}^{\;\nu}=\frac{i\kappa\epsilon(n+\nu+1-i\epsilon)(n+\nu+1+i\epsilon)^2}{(n+\nu+1)(2n+2\nu+3)},\\
			&\beta_{n}^{\;\nu}=(n+\nu)(n+\nu+1)-l(l+1)+\epsilon^2(\kappa+1)+\frac{\kappa  \epsilon ^4}{(n+\nu) (n+\nu+1)},\\
			&\gamma_{n}^{\;\nu}=-\frac{i \kappa  \epsilon  (n+\nu-i \epsilon )^2 (n+\nu+i \epsilon )}{(n+\nu) (2n+2\nu-1)}.
			\end{empheq}
		\end{subequations}
	It is useful to introduce
	\begin{equation}
	R_{n}(\nu)\equiv\frac{a_{n}^{\;\nu}}{a_{n-1}^{\;\nu}}\;\;\text{and}\;\;L_{n}(\nu)\equiv\frac{a_{n}^{\;\nu}}{a_{n+1}^{\;\nu}},
	\end{equation}
	they could also be expressed by continued fractions
	\begin{subequations}
		\begin{empheq}{align}
		&R_n(\nu)=-\frac{\gamma_{n}^{\;\nu}}{\beta_{n}^{\;\nu}+\alpha_{n}^{\;\nu}R_{n+1}(\nu)}=-\frac{\gamma_{n}^{\;\nu}}{\beta_{n}^{\;\nu}-}
		\frac{\alpha_{n}^{\;\nu}\gamma_{n+1}^{\;\nu}}{\beta_{n+1}^{\;\nu}-}
		\frac{\alpha_{n+1}^{\;\nu}\gamma_{n+2}^{\;\nu}}{\beta_{n+2}^{\;\nu}-}\cdots,\\
		&L_n(\nu)=-\frac{\alpha_{n}^{\;\nu}}{\beta_{n}^{\;\nu}+\gamma_{n}^{\;\nu}L_{n-1}(\nu)}=-\frac{\alpha_{n}^{\;\nu}}{\beta_{n}^{\;\nu}-}
		\frac{\alpha_{n-1}^{\;\nu}\gamma_{n}^{\;\nu}}{\beta_{n-1}^{\;\nu}-}
		\frac{\alpha_{n-2}^{\;\nu}\gamma_{n-1}^{\;\nu}}{\beta_{n-2}^{\;\nu}-}\cdots.
		\end{empheq}
	\end{subequations}
	
	The solution of the three-term recurrence relation is "minimal" (i.e., the series expansion in Eq.(\ref{Eq: series Hyper}) is convergent)
	if and only if the renormalized angular momentum $\nu$ satisfies
	\begin{equation}\label{Eq: root of nu}
	\beta_{n}^{\;\nu}+\alpha_{n}^{\;\nu}R_{n+1}(\nu)+\gamma_{n}^{\;\nu}L_{n-1}(\nu)=0.
	\end{equation}
	Note that the value of $n$ of this equation is arbitrary, it is convenient to set $n=0$ to solve for $\nu$.
	Note also that Eq.(\ref{Eq: root of nu}) contains an infinite number of roots:
	(i) if $\nu$ is a solution of Eq.(\ref{Eq: root of nu}), $\nu\pm1$ is also a solution;
	(ii) if $\nu$ is a solution, $-\nu-1$ is also a solution.
	However, not all of these solutions can be used to build up the radial solution we want.
	By definition, we must have $\nu\rightarrow l$ (or $\nu\rightarrow-l-1$) in the limit $\epsilon\rightarrow0$.
	Thus, we choose $\nu$ such that it approaches to $l$ as $\epsilon\rightarrow0$.
	
	Given the eigenvalue of $\nu$, it is straightforward to compute the series coefficients $a_{n}^{\;\nu}$
	from the three-term recurrence relation (\ref{Eq: series Hyper}). From Eqs.(\ref{Eq: alpha_n}),
	one finds $\alpha_{-n}^{-\nu-1}=\gamma_{n}^{\;\nu}$ and $\gamma_{-n}^{-\nu-1}=\alpha_{n}^{\;\nu}$
	so that $a_{-n}^{-\nu-1}$ obeys the same recursion relation as $a_{n}^{\;\nu}$ does.
	This implies that if we take $a_0^{\;\nu}=a_0^{-\nu-1}=1$, we have $a_{n}^{\;\nu}=a_{-n}^{-\nu-1}$.
	From Eq.(\ref{Eq: root of nu}), we have
	\begin{equation}\label{Eq: large n of an}
	\lim\limits_{n\rightarrow\infty}n\frac{a_{n}^{\;\nu}}{a_{n-1}^{\;\nu}}
	=-\lim\limits_{n\rightarrow-\infty}n\frac{a_{n}^{\;\nu}}{a_{n+1}^{\;\nu}}
	=\frac{i\kappa\epsilon}{2}.
	\end{equation}
	Combining the large $n$ behavior of hypergeometric functions, one finds
		\begin{equation}
		\lim\limits_{n\rightarrow\infty}\frac{na_{n}^{\;\nu}p_{n+\nu}}{a_{n-1}^{\;\nu}p_{n+\nu-1}}
		=-\lim\limits_{n\rightarrow-\infty}\frac{na_{n}^{\;\nu}p_{n+\nu}}{a_{n+1}^{\;\nu}p_{n+\nu+1}}
		=\frac{i\kappa\epsilon}{2}\left[1-2x+\left((1-2x)^2-1\right)^{1/2}\right].
		\end{equation}
	This implies that the series of hypergeometric functions (\ref{Eq: series Hyper}) converges in all over the complex plane of $x$ except for $|x|=\infty$.
	Thus, we have obtained the ingoing-wave radial solution which is valid for $|x|<\infty$.
	We note a property of the hypergeometric function
		\begin{equation}
		F(a,b;c;x)=\frac{\Gamma(c)\Gamma(b-a)}{\Gamma(c-a)\Gamma(b)}(1-x)^{-a}F\left(a,c-b;a+1-b;\frac{1}{1-x}\right)
		+\left(a\leftrightarrow b\right).
		\end{equation}
	This enables us to express the radial solution as
	\begin{equation}\label{Eq: near horizon solution}
	R^{\text{in}}=R_0^{\;\nu}+R_0^{-\nu-1},
	\end{equation}
	where
		\begin{equation}
		R_0^{\;\nu}=e^{i\kappa\epsilon x}(-x)^{-i\epsilon}(1-x)^{i\epsilon+\nu}
		\sum_{n=-\infty}^{\infty}a_{n}^{\;\nu}\frac{\Gamma(1-2i\epsilon)\Gamma(2n+2\nu+1)}{\Gamma(n+\nu+1-i\epsilon)^2}
		(1-x)^n F\left(-n-\nu-i\epsilon,-n-\nu-i\epsilon;-2n-2\nu;\frac{1}{1-x}\right).
		\end{equation}
	Here, we have used the property $a_{n}^{\;\nu}=a_{-n}^{-\nu-1}$. Clearly, Eq.(\ref{Eq: near horizon solution})
	explicitly exhibits the symmetry of $R^{\text{in}}$ under the interchange $\nu\leftrightarrow-\nu-1$.
	
	\subsection{Far region solution in series of Coulomb wave functions}
	The solution in the form of series of hypergeometric functions discussed in Appendix \ref{Subsec: near horizon} is convergent at any finite value of $r$.
	However, it does not converge at infinity. The analytic solution convergent at infinity was obtained by Leaver as a series of Coulomb functions \cite{PhysRevD.34.384}.
	Here, we follow the procedure in Ref.\cite{Mano:1996mf} to obtain the radial solution of Eq.(\ref{Eq: the radial eq}) that is convergent at infinity.
	
	First, we define a variable $z=\omega\left(r-Q^2/M\right)=\kappa\epsilon(1-x)$, and introduce the following form
	\begin{equation}\label{Eq: sol at inf}
	R_C=z^{-1}\left(1-\frac{\kappa\epsilon}{z}\right)^{-i\epsilon}f(z).
	\end{equation}
	Then the radial equation (\ref{Eq: the radial eq}) becomes
		\begin{equation}\label{Eq: eq of f(x) nu}
		z^2f''+\left[z^2+2\epsilon z-\nu(\nu+1)\right]f=\kappa\epsilon z\left(f''+f\right)+\kappa\epsilon(1-2i\epsilon)f'+\left[l(l+1)-\nu(\nu+1)-\epsilon^2\right]f,
		\end{equation}
	where a prime denote $d/dz$. As in Eq.(\ref{Eq: eq of p(x) nu}), we have introduced the renormalized angular momentum $\nu$ in this equation.
	We denote the solution specified by the index $\nu$ as $f_{\nu}(z)$, and expand it in terms of Coulomb wave functions as
	\begin{equation}\label{Eq: series Coulomb}
	f_{\nu}(z)=\sum_{n=-\infty}^{\infty}b_{n}^{\;\nu}F_{n+\nu}(z),
	\end{equation}
	where the Coulomb wave function is given by
		\begin{equation}\label{Eq: Coulomb}
		F_{n+\nu}(z)=e^{-iz}(2z)^{n+\nu}z\frac{\Gamma(n+\nu+1+i\epsilon)}{\Gamma(2n+2\nu+2)}\Phi(n+\nu+1+i\epsilon,2n+2\nu+2;2iz),
		\end{equation}
	and $\Phi$ denotes the regular confluent hypergeometric function. It can be shown that the Coulomb wave functions given in Eq.(\ref{Eq: Coulomb}) satisfy the following recurrence relations
		\begin{equation}
		\frac{1}{z}F_{n+\nu}=\frac{n-\tilde{a}+2\tilde{b}}{(\tilde{b}+n) (2\tilde{b}+2n-1)}F_{n+\nu+1}
		+\frac{i (\tilde{b}-\tilde{a})}{(\tilde{b}+n-1) (\tilde{b}+n)}F_{n+\nu}
		+\frac{\tilde{a}+n-1}{(\tilde{b}+n-1) (2 \tilde{b}+2 n-1)}F_{n+\nu-1},
		\end{equation}
		\begin{equation}
		F_{n+\nu}'=\frac{(\tilde{b}+n-1) (\tilde{a}-2 \tilde{b}-n)}{(\tilde{b}+n) (2\tilde{b}+2 n-1)}F_{n+\nu+1}
		+\frac{i (\tilde{b}-\tilde{a})}{(\tilde{b}+n-1) (\tilde{b}+n)}F_{n+\nu}
		+\frac{(\tilde{a}+n-1) (\tilde{b}+n)}{(\tilde{b}+n-1) (2 \tilde{b}+2 n-1)}F_{n+\nu-1},
		\end{equation}
	where $\tilde{a}=\nu+1+i\epsilon$ and $\tilde{b}=\nu+1$. Substituting Eq.(\ref{Eq: series Coulomb}) into Eq.(\ref{Eq: eq of f(x) nu})
	and using the above recurrence relations, one finds a three-term recurrence relation among the expansion coefficients $b_{n}^{\;\nu}$
	\begin{equation}\label{Eq: recurrence bn}
	\tilde{\alpha}_{n}^{\;\nu}b_{n+1}^{\;\nu}+\tilde{\beta}_{n}^{\;\nu}b_{n}^{\;\nu}+\tilde{\gamma}_{n}^{\;\nu}b_{n-1}^{\;\nu}=0,
	\end{equation}
	with
	\begin{subequations}\label{Eq: tidal alpha_n}
		\begin{empheq}{align}
		&\tilde{\alpha}_{n}^{\;\nu}=i\frac{(n+\nu+1+i\epsilon)}{(n+\nu+1-i\epsilon)}\alpha_{n}^{\;\nu},\\
		&\tilde{\beta}_{n}^{\;\nu}=\beta_{n}^{\;\nu},\\
		&\tilde{\gamma}_{n}^{\;\nu}=-i\frac{(n+\nu-i\epsilon)}{(n+\nu+i\epsilon)}\gamma_{n}^{\;\nu},
		\end{empheq}
	\end{subequations}
	where $\alpha_{n}^{\;\nu}$, $\beta_{n}^{\;\nu}$ and $\gamma_{n}^{\;\nu}$ are defined in Eqs.(\ref{Eq: alpha_n}).
	Clearly, the recurrence relation (\ref{Eq: recurrence bn}) is transformed to the one in Eq.(\ref{Eq: recurrence an})
	if we redefine the coefficients as
	\begin{equation}\label{Eq: an to bn}
	b_{n}^{\;\nu}=(-i)^n\frac{\left(\nu+1-i\epsilon\right)_n}{(\nu+1+i\epsilon)_n}a_{n}^{\;\nu},
	\end{equation}
	where $(x)_n\equiv\Gamma(x+n)/\Gamma(x)$. Since the recurrence relation obtained for the Coulomb expansion case is identical to the one for the hypergeometric case,
	the renormalized angular momentum $\nu$ from both solutions are the same. This is crucial in matching these two solutions which we will discuss later. 	
	From Eqs.(\ref{Eq: large n of an}) and (\ref{Eq: an to bn}), we find
	\begin{equation}\label{Eq: large n of nn}
	\lim\limits_{n\rightarrow\infty}n\frac{b_{n}^{\;\nu}}{b_{n-1}^{\;\nu}}
	=\lim\limits_{n\rightarrow-\infty}n\frac{b_{n}^{\;\nu}}{b_{n+1}^{\;\nu}}
	=\frac{\kappa\epsilon}{2}.
	\end{equation}
	By combing the large $n$ behavior of the Coulomb wave functions, we have
	\begin{equation}
	\lim\limits_{n\rightarrow\infty}\frac{b_{n}^{\;\nu}F_{n+\nu}}{b_{n-1}^{\;\nu}F_{n+\nu-1}}
	=\lim\limits_{n\rightarrow-\infty}\frac{b_{n}^{\;\nu}F_{n+\nu}}{b_{n+1}^{\;\nu}F_{n+\nu+1}}
	=\frac{\kappa\epsilon}{z}.
	\end{equation}
	This implies that the series of Coulomb wave functions (\ref{Eq: series Coulomb}) converges at $z>\kappa\epsilon$ or equivalently $r>r_+$. By using the formula
		\begin{equation}
		\Phi(a,c;z)=\frac{\Gamma(c)}{\Gamma(c-a)}e^{ia\pi}\Psi(a,c;z)+\frac{\Gamma(c)}{\Gamma(a)}e^{i\pi(a-c)}e^{x}\Psi(c-a,c;-z),
		\end{equation}
	and Eq.(\ref{Eq: an to bn}), we can rewrite Eq.(\ref{Eq: sol at inf}) as
	\begin{equation}
	R_{C}^{\nu}=R_{C\;\text{in}}^{\nu}+R_{C\;\text{out}}^{\nu},
	\end{equation}
	with
		\begin{equation}
		\begin{aligned}
		R_{C\;\text{in}}^{\nu}=&2^{\nu}e^{i\pi(\nu+1+i\epsilon)}
		\frac{\Gamma (\nu +1 -i \epsilon)}{\Gamma (\nu +1 +i \epsilon)}
		e^{-i z}z^{\nu+i\epsilon}(z-\kappa \epsilon )^{-i \epsilon }\sum _{n=-\infty }^{\infty } i^n a_n^{\;\nu}(2 z)^n \Psi (n+\nu+1+i \epsilon,2n+2\nu+2;2 i z),
		\end{aligned}
		\end{equation}
		\begin{equation}
		\begin{aligned}
		R_{C\;\text{out}}^{\nu}=&2^{\nu}e^{-i\pi(\nu+1-i\epsilon)}
		e^{i z}z^{\nu+i\epsilon}(z-\kappa \epsilon )^{-i \epsilon }\sum _{n=-\infty }^{\infty } i^n\frac{\left(\nu+1-i\epsilon\right)_n}{(\nu+1+i\epsilon)_n} a_n^{\;\nu}(2 z)^n \Psi (n+\nu+1-i \epsilon,2n+2\nu+2;-2 i z),
		\end{aligned}
		\end{equation}
	where $R_{C\;\text{in}}^{\nu}$ and $R_{C\;\text{out}}^{\nu}$ are the ingoing wave and outgoing wave solutions at infinity, respectively. And $\Psi$ is the irregular confluent hypergeometric function.
	At large $|z|$, $\Psi$ has the following asymptotic behavior
	\begin{equation}
	\Psi(a,c;z)\rightarrow z^{-a}\;\;\;\text{as}\;|z|\rightarrow\infty.
	\end{equation}
	Thus, at infinity, $R_{C\;\text{in}}^{\nu}$ and $R_{C\;\text{out}}^{\nu}$ behave as
	\begin{equation}
	R_{C\;\text{in}}^{\nu}=A^{\nu}_{\text{in}}z^{-1}e^{-i(z+\epsilon\ln z)},
	\end{equation}
	\begin{equation}
	R_{C\;\text{out}}^{\nu}=A^{\nu}_{\text{out}}z^{-1}e^{i(z+\epsilon\ln z)},
	\end{equation}
	where
	\begin{equation}
	A^{\nu}_{\text{in}}=2^{-1-i\epsilon}e^{i\frac{\pi}{2}(\nu+1+i\epsilon)}\frac{\Gamma(1+\nu-i\epsilon)}{\Gamma(1+\nu+i\epsilon)}\sum_{n=-\infty}^{\infty}a_{n}^{\;\nu},
	\end{equation}
	\begin{equation}
	A^{\nu}_{\text{out}}=2^{-1+i\epsilon}e^{-i\frac{\pi}{2}(\nu+1-i\epsilon)}\sum_{n=-\infty}^{\infty}(-1)^n\frac{\left(\nu+1-i\epsilon\right)_n}{(\nu+1+i\epsilon)_n}a_n^{\;\nu},
	\end{equation}
	Note that there exist another independent solution which is obtained by replacing $\nu$ with $-\nu-1$. It is given by
	\begin{equation}
	R_{C}^{-\nu-1}=-ie^{-i\pi\nu}\frac{\sin\pi(\nu+i\epsilon)}{\sin\pi(\nu-i\epsilon)}R_{C\;\text{in}}^{\nu}+ie^{i\pi\nu}R_{C\;\text{out}}^{\nu}.
	\end{equation}
	\subsection{Matching of the two solutions}
	We have obtained two types of solutions $R_0^{\;\nu}$ and $R_C^{\;\nu}$ of Eq.(\ref{Eq: the radial eq}).
	Note that both of them could be expressed in terms of $z=\kappa\epsilon(1-x)$,
	and they are convergent for a wide range of $z$, i.e., $\kappa\epsilon<z<\infty$.
	Using the large $z$ behavior of the (confluent) hypergeometric functions, one finds that $R_0^{\;\nu}$ must be proportional to $R_C^{\;\nu}$, i.e.,
	\begin{equation}
	R_0^{\;\nu}=K_{\nu}R_C^{\;\nu}.
	\end{equation}
	The constant factor $K_{\nu}$ is obtained by comparing the asymptotic behavior of $R_0^{\;\nu}$ and $R_C^{\;\nu}$. It is given by
		\begin{equation}
		\begin{aligned}	
		K_{\nu}=&e^{i\kappa\epsilon}(2\kappa\epsilon)^{-\nu-p}i^p\frac{\Gamma(1-2i\epsilon)\Gamma(p+2\nu+2)}{\Gamma(p+\nu+1+i\epsilon)^3}
		\times\left(\sum_{n=p}^{\infty}(-1)^n\frac{\Gamma(n+p+2\nu+1)}{(n-p)!}\frac{\Gamma(n+\nu+1+i\epsilon)^2}{\Gamma(n+\nu+1-i\epsilon)^2}a_n^{\;\nu}\right)\\
		&\times\left(\sum_{n=-\infty}^{p}\frac{(-1)^n}{(p+2\nu+2)_n(p-n)!}\frac{\left(\nu+1-i\epsilon\right)_n}{(\nu+1+i\epsilon)_n}a_n^{\;\nu}\right)^{-1},
		\end{aligned}
		\end{equation}
	where $p$ is an arbitrary integer and $K_{\nu}$ is independent of the choice of $p$. This can be used as a check of the calculation.
	Thus, the ingoing wave solution (\ref{Eq: ingoing soln}) could also be expressed the Coulomb functions, i.e., 	
		\begin{equation}\label{Eq: sol via MST}
		R^{\text{in}}=K_{\nu}R_C^{\;\nu}+K_{-\nu-1}R_C^{\;-\nu-1}
		=\left(K_{\nu}-ie^{-i\pi\nu}\frac{\sin\pi(\nu+i\epsilon)}{\sin\pi(\nu-i\epsilon)}K_{-\nu-1}\right)R_{C\;\text{in}}^{\nu}
		+\left(K_{\nu}+ie^{i\pi\nu}K_{-\nu-1}\right)R_{C\;\text{out}}^{\nu}.
		\end{equation}
	From Eq.(\ref{Eq: bcs infty}), the radial function has the form
	\begin{equation}
	R(r)\sim A^{(-)}_{\lambda-1/2}\left(\frac{r}{2M}\right)^{-1-i2M\omega}e^{-i\omega r}+A^{(+)}_{\lambda-1/2}\times(\omega\rightarrow-\omega)
	\end{equation}	
	as $r\rightarrow\infty$. Here, we have used $l=\lambda-1/2$.
	Finally, comparing this equation with Eq.(\ref{Eq: sol via MST}) and using $z=\omega(r-Q^2/M)$, we obtain
	\begin{equation}
	A^{(-)}_{\lambda-1/2}=\epsilon^{-1}e^{-i\epsilon\ln\epsilon}\left(K_{\nu}-ie^{-i\pi\nu}\frac{\sin\pi(\nu+i\epsilon)}{\sin\pi(\nu-i\epsilon)}K_{-\nu-1}\right)A_{\text{in}}^{\nu},
	\end{equation}
	\begin{equation}
	A^{(+)}_{\lambda-1/2}=\epsilon^{-1}e^{i\epsilon\ln\epsilon}\left(K_{\nu}+ie^{i\pi\nu}K_{-\nu-1}\right)A_{\text{out}}^{\nu}.
	\end{equation}
\end{widetext}

\bibliography{crosssection}

\begin{thebibliography}{57}%
\makeatletter
\providecommand \@ifxundefined [1]{%
 \@ifx{#1\undefined}
}%
\providecommand \@ifnum [1]{%
 \ifnum #1\expandafter \@firstoftwo
 \else \expandafter \@secondoftwo
 \fi
}%
\providecommand \@ifx [1]{%
 \ifx #1\expandafter \@firstoftwo
 \else \expandafter \@secondoftwo
 \fi
}%
\providecommand \natexlab [1]{#1}%
\providecommand \enquote  [1]{``#1''}%
\providecommand \bibnamefont  [1]{#1}%
\providecommand \bibfnamefont [1]{#1}%
\providecommand \citenamefont [1]{#1}%
\providecommand \href@noop [0]{\@secondoftwo}%
\providecommand \href [0]{\begingroup \@sanitize@url \@href}%
\providecommand \@href[1]{\@@startlink{#1}\@@href}%
\providecommand \@@href[1]{\endgroup#1\@@endlink}%
\providecommand \@sanitize@url [0]{\catcode `\\12\catcode `\$12\catcode
  `\&12\catcode `\#12\catcode `\^12\catcode `\_12\catcode `\%12\relax}%
\providecommand \@@startlink[1]{}%
\providecommand \@@endlink[0]{}%
\providecommand \url  [0]{\begingroup\@sanitize@url \@url }%
\providecommand \@url [1]{\endgroup\@href {#1}{\urlprefix }}%
\providecommand \urlprefix  [0]{URL }%
\providecommand \Eprint [0]{\href }%
\providecommand \doibase [0]{http://dx.doi.org/}%
\providecommand \selectlanguage [0]{\@gobble}%
\providecommand \bibinfo  [0]{\@secondoftwo}%
\providecommand \bibfield  [0]{\@secondoftwo}%
\providecommand \translation [1]{[#1]}%
\providecommand \BibitemOpen [0]{}%
\providecommand \bibitemStop [0]{}%
\providecommand \bibitemNoStop [0]{.\EOS\space}%
\providecommand \EOS [0]{\spacefactor3000\relax}%
\providecommand \BibitemShut  [1]{\csname bibitem#1\endcsname}%
\let\auto@bib@innerbib\@empty
\bibitem [{\citenamefont {Abbott}\ \emph {et~al.}(2016)\citenamefont {Abbott}
  \emph {et~al.}}]{Abbott:2016blz}%
  \BibitemOpen
  \bibfield  {author} {\bibinfo {author} {\bibfnamefont {B.}~\bibnamefont
  {Abbott}} \emph {et~al.} (\bibinfo {collaboration} {LIGO Scientific,
  Virgo}),\ }\href {\doibase 10.1103/PhysRevLett.116.061102} {\bibfield
  {journal} {\bibinfo  {journal} {Phys. Rev. Lett.}\ }\textbf {\bibinfo
  {volume} {116}},\ \bibinfo {pages} {061102} (\bibinfo {year} {2016})},\
  \Eprint {http://arxiv.org/abs/1602.03837} {arXiv:1602.03837 [gr-qc]}
  \BibitemShut {NoStop}%
\bibitem [{Gra()}]{GraceDB}%
  \BibitemOpen
  \href@noop {} {}\bibinfo {howpublished}
  {\url{https://gracedb.ligo.org/latest}}\BibitemShut {NoStop}%
\bibitem [{EHT()}]{EHT}%
  \BibitemOpen
  \href@noop {} {}\bibinfo {howpublished}
  {\url{https://eventhorizontelescope.org}}\BibitemShut {NoStop}%
\bibitem [{\citenamefont {Matzner}(1968)}]{doi:10.1063/1.1664470}%
  \BibitemOpen
  \bibfield  {author} {\bibinfo {author} {\bibfnamefont {R.~A.}\ \bibnamefont
  {Matzner}},\ }\href {\doibase 10.1063/1.1664470} {\bibfield  {journal}
  {\bibinfo  {journal} {Journal of Mathematical Physics}\ }\textbf {\bibinfo
  {volume} {9}},\ \bibinfo {pages} {163} (\bibinfo {year} {1968})},\ \Eprint
  {http://arxiv.org/abs/https://doi.org/10.1063/1.1664470}
  {https://doi.org/10.1063/1.1664470} \BibitemShut {NoStop}%
\bibitem [{\citenamefont {Futterman}\ \emph {et~al.}(1988)\citenamefont
  {Futterman}, \citenamefont {Handler},\ and\ \citenamefont
  {Matzner}}]{futterman_handler_matzner_1988}%
  \BibitemOpen
  \bibfield  {author} {\bibinfo {author} {\bibfnamefont {J.~A.~H.}\
  \bibnamefont {Futterman}}, \bibinfo {author} {\bibfnamefont {F.~A.}\
  \bibnamefont {Handler}}, \ and\ \bibinfo {author} {\bibfnamefont {R.~A.}\
  \bibnamefont {Matzner}},\ }\href {\doibase 10.1017/CBO9780511735615} {\emph
  {\bibinfo {title} {Scattering from Black Holes}}},\ Cambridge Monographs on
  Mathematical Physics\ (\bibinfo  {publisher} {Cambridge University Press},\
  \bibinfo {year} {1988})\BibitemShut {NoStop}%
\bibitem [{\citenamefont {Matzner}\ \emph {et~al.}(1985)\citenamefont
  {Matzner}, \citenamefont {DeWitte-Morette}, \citenamefont {Nelson},\ and\
  \citenamefont {Zhang}}]{PhysRevD.31.1869}%
  \BibitemOpen
  \bibfield  {author} {\bibinfo {author} {\bibfnamefont {R.~A.}\ \bibnamefont
  {Matzner}}, \bibinfo {author} {\bibfnamefont {C.}~\bibnamefont
  {DeWitte-Morette}}, \bibinfo {author} {\bibfnamefont {B.}~\bibnamefont
  {Nelson}}, \ and\ \bibinfo {author} {\bibfnamefont {T.-R.}\ \bibnamefont
  {Zhang}},\ }\href {\doibase 10.1103/PhysRevD.31.1869} {\bibfield  {journal}
  {\bibinfo  {journal} {Phys. Rev. D}\ }\textbf {\bibinfo {volume} {31}},\
  \bibinfo {pages} {1869} (\bibinfo {year} {1985})}\BibitemShut {NoStop}%
\bibitem [{\citenamefont {Handler}\ and\ \citenamefont
  {Matzner}(1980)}]{PhysRevD.22.2331}%
  \BibitemOpen
  \bibfield  {author} {\bibinfo {author} {\bibfnamefont {F.~A.}\ \bibnamefont
  {Handler}}\ and\ \bibinfo {author} {\bibfnamefont {R.~A.}\ \bibnamefont
  {Matzner}},\ }\href {\doibase 10.1103/PhysRevD.22.2331} {\bibfield  {journal}
  {\bibinfo  {journal} {Phys. Rev. D}\ }\textbf {\bibinfo {volume} {22}},\
  \bibinfo {pages} {2331} (\bibinfo {year} {1980})}\BibitemShut {NoStop}%
\bibitem [{\citenamefont {Leite}\ \emph
  {et~al.}(2019{\natexlab{a}})\citenamefont {Leite}, \citenamefont {Dolan},\
  and\ \citenamefont {Crispino}}]{PhysRevD.100.084025}%
  \BibitemOpen
  \bibfield  {author} {\bibinfo {author} {\bibfnamefont {L.~C.~S.}\
  \bibnamefont {Leite}}, \bibinfo {author} {\bibfnamefont {S.~R.}\ \bibnamefont
  {Dolan}}, \ and\ \bibinfo {author} {\bibfnamefont {L.~C.~B.}\ \bibnamefont
  {Crispino}},\ }\href {\doibase 10.1103/PhysRevD.100.084025} {\bibfield
  {journal} {\bibinfo  {journal} {Phys. Rev. D}\ }\textbf {\bibinfo {volume}
  {100}},\ \bibinfo {pages} {084025} (\bibinfo {year}
  {2019}{\natexlab{a}})}\BibitemShut {NoStop}%
\bibitem [{\citenamefont {Mashhoon}(1974)}]{PhysRevD.10.1059}%
  \BibitemOpen
  \bibfield  {author} {\bibinfo {author} {\bibfnamefont {B.}~\bibnamefont
  {Mashhoon}},\ }\href {\doibase 10.1103/PhysRevD.10.1059} {\bibfield
  {journal} {\bibinfo  {journal} {Phys. Rev. D}\ }\textbf {\bibinfo {volume}
  {10}},\ \bibinfo {pages} {1059} (\bibinfo {year} {1974})}\BibitemShut
  {NoStop}%
\bibitem [{\citenamefont {Glampedakis}\ and\ \citenamefont
  {Andersson}(2001)}]{Glampedakis:2001cx}%
  \BibitemOpen
  \bibfield  {author} {\bibinfo {author} {\bibfnamefont {K.}~\bibnamefont
  {Glampedakis}}\ and\ \bibinfo {author} {\bibfnamefont {N.}~\bibnamefont
  {Andersson}},\ }\href {\doibase 10.1088/0264-9381/18/10/309} {\bibfield
  {journal} {\bibinfo  {journal} {Class. Quant. Grav.}\ }\textbf {\bibinfo
  {volume} {18}},\ \bibinfo {pages} {1939} (\bibinfo {year} {2001})},\ \Eprint
  {http://arxiv.org/abs/gr-qc/0102100} {arXiv:gr-qc/0102100} \BibitemShut
  {NoStop}%
\bibitem [{\citenamefont {S\'anchez}(1978)}]{PhysRevD.18.1798}%
  \BibitemOpen
  \bibfield  {author} {\bibinfo {author} {\bibfnamefont {N.}~\bibnamefont
  {S\'anchez}},\ }\href {\doibase 10.1103/PhysRevD.18.1798} {\bibfield
  {journal} {\bibinfo  {journal} {Phys. Rev. D}\ }\textbf {\bibinfo {volume}
  {18}},\ \bibinfo {pages} {1798} (\bibinfo {year} {1978})}\BibitemShut
  {NoStop}%
\bibitem [{\citenamefont {Dolan}\ \emph {et~al.}(2006)\citenamefont {Dolan},
  \citenamefont {Doran},\ and\ \citenamefont {Lasenby}}]{PhysRevD.74.064005}%
  \BibitemOpen
  \bibfield  {author} {\bibinfo {author} {\bibfnamefont {S.}~\bibnamefont
  {Dolan}}, \bibinfo {author} {\bibfnamefont {C.}~\bibnamefont {Doran}}, \ and\
  \bibinfo {author} {\bibfnamefont {A.}~\bibnamefont {Lasenby}},\ }\href
  {\doibase 10.1103/PhysRevD.74.064005} {\bibfield  {journal} {\bibinfo
  {journal} {Phys. Rev. D}\ }\textbf {\bibinfo {volume} {74}},\ \bibinfo
  {pages} {064005} (\bibinfo {year} {2006})}\BibitemShut {NoStop}%
\bibitem [{\citenamefont {Crispino}\ \emph
  {et~al.}(2009{\natexlab{a}})\citenamefont {Crispino}, \citenamefont {Dolan},\
  and\ \citenamefont {Oliveira}}]{PhysRevLett.102.231103}%
  \BibitemOpen
  \bibfield  {author} {\bibinfo {author} {\bibfnamefont {L.~C.~B.}\
  \bibnamefont {Crispino}}, \bibinfo {author} {\bibfnamefont {S.~R.}\
  \bibnamefont {Dolan}}, \ and\ \bibinfo {author} {\bibfnamefont {E.~S.}\
  \bibnamefont {Oliveira}},\ }\href {\doibase 10.1103/PhysRevLett.102.231103}
  {\bibfield  {journal} {\bibinfo  {journal} {Phys. Rev. Lett.}\ }\textbf
  {\bibinfo {volume} {102}},\ \bibinfo {pages} {231103} (\bibinfo {year}
  {2009}{\natexlab{a}})}\BibitemShut {NoStop}%
\bibitem [{\citenamefont {Dolan}(2008)}]{Dolan:2008kf}%
  \BibitemOpen
  \bibfield  {author} {\bibinfo {author} {\bibfnamefont {S.~R.}\ \bibnamefont
  {Dolan}},\ }\href {\doibase 10.1088/0264-9381/25/23/235002} {\bibfield
  {journal} {\bibinfo  {journal} {Class. Quant. Grav.}\ }\textbf {\bibinfo
  {volume} {25}},\ \bibinfo {pages} {235002} (\bibinfo {year} {2008})},\
  \Eprint {http://arxiv.org/abs/0801.3805} {arXiv:0801.3805 [gr-qc]}
  \BibitemShut {NoStop}%
\bibitem [{\citenamefont {Leite}\ \emph
  {et~al.}(2019{\natexlab{b}})\citenamefont {Leite}, \citenamefont {Benone},\
  and\ \citenamefont {Crispino}}]{Leite:2019eis}%
  \BibitemOpen
  \bibfield  {author} {\bibinfo {author} {\bibfnamefont {L.~C.}\ \bibnamefont
  {Leite}}, \bibinfo {author} {\bibfnamefont {C.~L.}\ \bibnamefont {Benone}}, \
  and\ \bibinfo {author} {\bibfnamefont {L.~C.}\ \bibnamefont {Crispino}},\
  }\href {\doibase 10.1016/j.physletb.2019.06.027} {\bibfield  {journal}
  {\bibinfo  {journal} {Phys. Lett. B}\ }\textbf {\bibinfo {volume} {795}},\
  \bibinfo {pages} {496} (\bibinfo {year} {2019}{\natexlab{b}})},\ \Eprint
  {http://arxiv.org/abs/1907.04746} {arXiv:1907.04746 [gr-qc]} \BibitemShut
  {NoStop}%
\bibitem [{\citenamefont {Cotaescu}\ \emph {et~al.}(2016)\citenamefont
  {Cotaescu}, \citenamefont {Crucean},\ and\ \citenamefont
  {Sporea}}]{Cotaescu:2014jca}%
  \BibitemOpen
  \bibfield  {author} {\bibinfo {author} {\bibfnamefont {I.~I.}\ \bibnamefont
  {Cotaescu}}, \bibinfo {author} {\bibfnamefont {C.}~\bibnamefont {Crucean}}, \
  and\ \bibinfo {author} {\bibfnamefont {C.~A.}\ \bibnamefont {Sporea}},\
  }\href {\doibase 10.1140/epjc/s10052-016-3936-9} {\bibfield  {journal}
  {\bibinfo  {journal} {Eur. Phys. J. C}\ }\textbf {\bibinfo {volume} {76}},\
  \bibinfo {pages} {102} (\bibinfo {year} {2016})},\ \Eprint
  {http://arxiv.org/abs/1409.7201} {arXiv:1409.7201 [gr-qc]} \BibitemShut
  {NoStop}%
\bibitem [{\citenamefont {Macedo}\ \emph {et~al.}(2015)\citenamefont {Macedo},
  \citenamefont {de~Oliveira},\ and\ \citenamefont
  {Crispino}}]{PhysRevD.92.024012}%
  \BibitemOpen
  \bibfield  {author} {\bibinfo {author} {\bibfnamefont {C.~F.~B.}\
  \bibnamefont {Macedo}}, \bibinfo {author} {\bibfnamefont {E.~S.}\
  \bibnamefont {de~Oliveira}}, \ and\ \bibinfo {author} {\bibfnamefont
  {L.~C.~B.}\ \bibnamefont {Crispino}},\ }\href {\doibase
  10.1103/PhysRevD.92.024012} {\bibfield  {journal} {\bibinfo  {journal} {Phys.
  Rev. D}\ }\textbf {\bibinfo {volume} {92}},\ \bibinfo {pages} {024012}
  (\bibinfo {year} {2015})}\BibitemShut {NoStop}%
\bibitem [{\citenamefont {de~Oliveira}(2018)}]{deOliveira:2018kcq}%
  \BibitemOpen
  \bibfield  {author} {\bibinfo {author} {\bibfnamefont {E.~S.}\ \bibnamefont
  {de~Oliveira}},\ }\href {\doibase 10.1140/epjc/s10052-018-6316-9} {\bibfield
  {journal} {\bibinfo  {journal} {Eur. Phys. J.}\ }\textbf {\bibinfo {volume}
  {C78}},\ \bibinfo {pages} {876} (\bibinfo {year} {2018})},\ \Eprint
  {http://arxiv.org/abs/1805.04987} {arXiv:1805.04987 [gr-qc]} \BibitemShut
  {NoStop}%
\bibitem [{\citenamefont {Huang}\ \emph {et~al.}(2014)\citenamefont {Huang},
  \citenamefont {Liao}, \citenamefont {Chen},\ and\ \citenamefont
  {Wang}}]{Huang:2014nka}%
  \BibitemOpen
  \bibfield  {author} {\bibinfo {author} {\bibfnamefont {H.}~\bibnamefont
  {Huang}}, \bibinfo {author} {\bibfnamefont {P.}~\bibnamefont {Liao}},
  \bibinfo {author} {\bibfnamefont {J.}~\bibnamefont {Chen}}, \ and\ \bibinfo
  {author} {\bibfnamefont {Y.}~\bibnamefont {Wang}},\ }\href {\doibase
  10.1155/2014/231727} {\bibfield  {journal} {\bibinfo  {journal} {J. Grav.}\
  }\textbf {\bibinfo {volume} {2014}},\ \bibinfo {pages} {231727} (\bibinfo
  {year} {2014})}\BibitemShut {NoStop}%
\bibitem [{\citenamefont {Gußmann}(2017)}]{Gussmann:2016mkp}%
  \BibitemOpen
  \bibfield  {author} {\bibinfo {author} {\bibfnamefont {A.}~\bibnamefont
  {Gußmann}},\ }\href {\doibase 10.1088/1361-6382/aa606c} {\bibfield
  {journal} {\bibinfo  {journal} {Class. Quant. Grav.}\ }\textbf {\bibinfo
  {volume} {34}},\ \bibinfo {pages} {065007} (\bibinfo {year} {2017})},\
  \Eprint {http://arxiv.org/abs/1608.00552} {arXiv:1608.00552 [hep-th]}
  \BibitemShut {NoStop}%
\bibitem [{\citenamefont {Lin}\ \emph {et~al.}(2020{\natexlab{a}})\citenamefont
  {Lin}, \citenamefont {Jiang}, \citenamefont {Huang},\ and\ \citenamefont
  {Zhai}}]{Lin:2020rvv}%
  \BibitemOpen
  \bibfield  {author} {\bibinfo {author} {\bibfnamefont {R.-H.}\ \bibnamefont
  {Lin}}, \bibinfo {author} {\bibfnamefont {R.}~\bibnamefont {Jiang}}, \bibinfo
  {author} {\bibfnamefont {Y.}~\bibnamefont {Huang}}, \ and\ \bibinfo {author}
  {\bibfnamefont {X.-H.}\ \bibnamefont {Zhai}},\ }\href@noop {} {\  (\bibinfo
  {year} {2020}{\natexlab{a}})},\ \Eprint {http://arxiv.org/abs/2001.01363}
  {arXiv:2001.01363 [gr-qc]} \BibitemShut {NoStop}%
\bibitem [{\citenamefont {Ould El~Hadj}\ \emph {et~al.}(2020)\citenamefont
  {Ould El~Hadj}, \citenamefont {Stratton},\ and\ \citenamefont
  {Dolan}}]{OuldElHadj:2019kji}%
  \BibitemOpen
  \bibfield  {author} {\bibinfo {author} {\bibfnamefont {M.}~\bibnamefont {Ould
  El~Hadj}}, \bibinfo {author} {\bibfnamefont {T.}~\bibnamefont {Stratton}}, \
  and\ \bibinfo {author} {\bibfnamefont {S.~R.}\ \bibnamefont {Dolan}},\ }\href
  {\doibase 10.1103/PhysRevD.101.104035} {\bibfield  {journal} {\bibinfo
  {journal} {Phys. Rev. D}\ }\textbf {\bibinfo {volume} {101}},\ \bibinfo
  {pages} {104035} (\bibinfo {year} {2020})},\ \Eprint
  {http://arxiv.org/abs/1912.11348} {arXiv:1912.11348 [gr-qc]} \BibitemShut
  {NoStop}%
\bibitem [{\citenamefont {Stratton}\ and\ \citenamefont
  {Dolan}(2019)}]{Stratton:2019deq}%
  \BibitemOpen
  \bibfield  {author} {\bibinfo {author} {\bibfnamefont {T.}~\bibnamefont
  {Stratton}}\ and\ \bibinfo {author} {\bibfnamefont {S.~R.}\ \bibnamefont
  {Dolan}},\ }\href {\doibase 10.1103/PhysRevD.100.024007} {\bibfield
  {journal} {\bibinfo  {journal} {Phys. Rev. D}\ }\textbf {\bibinfo {volume}
  {100}},\ \bibinfo {pages} {024007} (\bibinfo {year} {2019})},\ \Eprint
  {http://arxiv.org/abs/1903.00025} {arXiv:1903.00025 [gr-qc]} \BibitemShut
  {NoStop}%
\bibitem [{\citenamefont {Dolan}\ and\ \citenamefont
  {Stratton}(2017)}]{Dolan:2017rtj}%
  \BibitemOpen
  \bibfield  {author} {\bibinfo {author} {\bibfnamefont {S.~R.}\ \bibnamefont
  {Dolan}}\ and\ \bibinfo {author} {\bibfnamefont {T.}~\bibnamefont
  {Stratton}},\ }\href {\doibase 10.1103/PhysRevD.95.124055} {\bibfield
  {journal} {\bibinfo  {journal} {Phys. Rev. D}\ }\textbf {\bibinfo {volume}
  {95}},\ \bibinfo {pages} {124055} (\bibinfo {year} {2017})},\ \Eprint
  {http://arxiv.org/abs/1702.06127} {arXiv:1702.06127 [gr-qc]} \BibitemShut
  {NoStop}%
\bibitem [{\citenamefont {Cotaescu}\ and\ \citenamefont
  {Sporea}(2019)}]{Cotaescu:2018etx}%
  \BibitemOpen
  \bibfield  {author} {\bibinfo {author} {\bibfnamefont {I.~I.}\ \bibnamefont
  {Cotaescu}}\ and\ \bibinfo {author} {\bibfnamefont {C.~A.}\ \bibnamefont
  {Sporea}},\ }\href {\doibase 10.1140/epjc/s10052-018-6525-2} {\bibfield
  {journal} {\bibinfo  {journal} {Eur. Phys. J. C}\ }\textbf {\bibinfo {volume}
  {79}},\ \bibinfo {pages} {15} (\bibinfo {year} {2019})},\ \Eprint
  {http://arxiv.org/abs/1811.07723} {arXiv:1811.07723 [gr-qc]} \BibitemShut
  {NoStop}%
\bibitem [{\citenamefont {Tominaga}\ \emph {et~al.}(1999)\citenamefont
  {Tominaga}, \citenamefont {Saijo},\ and\ \citenamefont
  {Maeda}}]{Tominaga:1999iy}%
  \BibitemOpen
  \bibfield  {author} {\bibinfo {author} {\bibfnamefont {K.}~\bibnamefont
  {Tominaga}}, \bibinfo {author} {\bibfnamefont {M.}~\bibnamefont {Saijo}}, \
  and\ \bibinfo {author} {\bibfnamefont {K.-i.}\ \bibnamefont {Maeda}},\ }\href
  {\doibase 10.1103/PhysRevD.60.024004} {\bibfield  {journal} {\bibinfo
  {journal} {Phys. Rev. D}\ }\textbf {\bibinfo {volume} {60}},\ \bibinfo
  {pages} {024004} (\bibinfo {year} {1999})},\ \Eprint
  {http://arxiv.org/abs/gr-qc/9901040} {arXiv:gr-qc/9901040} \BibitemShut
  {NoStop}%
\bibitem [{\citenamefont {Tominaga}\ \emph {et~al.}(2001)\citenamefont
  {Tominaga}, \citenamefont {Saijo},\ and\ \citenamefont
  {Maeda}}]{Tominaga:2000cs}%
  \BibitemOpen
  \bibfield  {author} {\bibinfo {author} {\bibfnamefont {K.}~\bibnamefont
  {Tominaga}}, \bibinfo {author} {\bibfnamefont {M.}~\bibnamefont {Saijo}}, \
  and\ \bibinfo {author} {\bibfnamefont {K.-i.}\ \bibnamefont {Maeda}},\ }\href
  {\doibase 10.1103/PhysRevD.63.124012} {\bibfield  {journal} {\bibinfo
  {journal} {Phys. Rev. D}\ }\textbf {\bibinfo {volume} {63}},\ \bibinfo
  {pages} {124012} (\bibinfo {year} {2001})},\ \Eprint
  {http://arxiv.org/abs/gr-qc/0009055} {arXiv:gr-qc/0009055} \BibitemShut
  {NoStop}%
\bibitem [{\citenamefont {Yennie}\ \emph {et~al.}(1954)\citenamefont {Yennie},
  \citenamefont {Ravenhall},\ and\ \citenamefont {Wilson}}]{PhysRev.95.500}%
  \BibitemOpen
  \bibfield  {author} {\bibinfo {author} {\bibfnamefont {D.~R.}\ \bibnamefont
  {Yennie}}, \bibinfo {author} {\bibfnamefont {D.~G.}\ \bibnamefont
  {Ravenhall}}, \ and\ \bibinfo {author} {\bibfnamefont {R.~N.}\ \bibnamefont
  {Wilson}},\ }\href {\doibase 10.1103/PhysRev.95.500} {\bibfield  {journal}
  {\bibinfo  {journal} {Phys. Rev.}\ }\textbf {\bibinfo {volume} {95}},\
  \bibinfo {pages} {500} (\bibinfo {year} {1954})}\BibitemShut {NoStop}%
\bibitem [{\citenamefont {Andersson}\ and\ \citenamefont
  {Thylwe}(1994)}]{Andersson_1994}%
  \BibitemOpen
  \bibfield  {author} {\bibinfo {author} {\bibfnamefont {N.}~\bibnamefont
  {Andersson}}\ and\ \bibinfo {author} {\bibfnamefont {K.-E.}\ \bibnamefont
  {Thylwe}},\ }\href {\doibase 10.1088/0264-9381/11/12/013} {\bibfield
  {journal} {\bibinfo  {journal} {Classical and Quantum Gravity}\ }\textbf
  {\bibinfo {volume} {11}},\ \bibinfo {pages} {2991} (\bibinfo {year}
  {1994})}\BibitemShut {NoStop}%
\bibitem [{\citenamefont {Andersson}(1994)}]{Andersson_1994_}%
  \BibitemOpen
  \bibfield  {author} {\bibinfo {author} {\bibfnamefont {N.}~\bibnamefont
  {Andersson}},\ }\href {\doibase 10.1088/0264-9381/11/12/014} {\bibfield
  {journal} {\bibinfo  {journal} {Classical and Quantum Gravity}\ }\textbf
  {\bibinfo {volume} {11}},\ \bibinfo {pages} {3003} (\bibinfo {year}
  {1994})}\BibitemShut {NoStop}%
\bibitem [{\citenamefont {Folacci}\ and\ \citenamefont {Ould
  El~Hadj}(2019)}]{Folacci:2019cmc}%
  \BibitemOpen
  \bibfield  {author} {\bibinfo {author} {\bibfnamefont {A.}~\bibnamefont
  {Folacci}}\ and\ \bibinfo {author} {\bibfnamefont {M.}~\bibnamefont {Ould
  El~Hadj}},\ }\href {\doibase 10.1103/PhysRevD.99.104079} {\bibfield
  {journal} {\bibinfo  {journal} {Phys. Rev. D}\ }\textbf {\bibinfo {volume}
  {99}},\ \bibinfo {pages} {104079} (\bibinfo {year} {2019})},\ \Eprint
  {http://arxiv.org/abs/1901.03965} {arXiv:1901.03965 [gr-qc]} \BibitemShut
  {NoStop}%
\bibitem [{\citenamefont {Folacci}\ and\ \citenamefont
  {Hadj}(2019)}]{PhysRevD.100.064009}%
  \BibitemOpen
  \bibfield  {author} {\bibinfo {author} {\bibfnamefont {A.}~\bibnamefont
  {Folacci}}\ and\ \bibinfo {author} {\bibfnamefont {M.~O.~E.}\ \bibnamefont
  {Hadj}},\ }\href {\doibase 10.1103/PhysRevD.100.064009} {\bibfield  {journal}
  {\bibinfo  {journal} {Phys. Rev. D}\ }\textbf {\bibinfo {volume} {100}},\
  \bibinfo {pages} {064009} (\bibinfo {year} {2019})}\BibitemShut {NoStop}%
\bibitem [{\citenamefont {Gibbons}\ and\ \citenamefont
  {Maeda}(1988)}]{Gibbons:1987ps}%
  \BibitemOpen
  \bibfield  {author} {\bibinfo {author} {\bibfnamefont {G.~W.}\ \bibnamefont
  {Gibbons}}\ and\ \bibinfo {author} {\bibfnamefont {K.-i.}\ \bibnamefont
  {Maeda}},\ }\href {\doibase 10.1016/0550-3213(88)90006-5} {\bibfield
  {journal} {\bibinfo  {journal} {Nucl. Phys.}\ }\textbf {\bibinfo {volume}
  {B298}},\ \bibinfo {pages} {741} (\bibinfo {year} {1988})}\BibitemShut
  {NoStop}%
\bibitem [{\citenamefont {Garfinkle}\ \emph {et~al.}(1991)\citenamefont
  {Garfinkle}, \citenamefont {Horowitz},\ and\ \citenamefont
  {Strominger}}]{Garfinkle:1990qj}%
  \BibitemOpen
  \bibfield  {author} {\bibinfo {author} {\bibfnamefont {D.}~\bibnamefont
  {Garfinkle}}, \bibinfo {author} {\bibfnamefont {G.~T.}\ \bibnamefont
  {Horowitz}}, \ and\ \bibinfo {author} {\bibfnamefont {A.}~\bibnamefont
  {Strominger}},\ }\href {\doibase 10.1103/PhysRevD.43.3140,
  10.1103/PhysRevD.45.3888} {\bibfield  {journal} {\bibinfo  {journal} {Phys.
  Rev.}\ }\textbf {\bibinfo {volume} {D43}},\ \bibinfo {pages} {3140} (\bibinfo
  {year} {1991})},\ \bibinfo {note} {[Erratum: Phys.
  Rev.D45,3888(1992)]}\BibitemShut {NoStop}%
\bibitem [{\citenamefont {Blaga}(2015)}]{Blaga:2014spa}%
  \BibitemOpen
  \bibfield  {author} {\bibinfo {author} {\bibfnamefont {C.}~\bibnamefont
  {Blaga}},\ }\href {\doibase 10.2298/SAJ1590041B} {\bibfield  {journal}
  {\bibinfo  {journal} {Serb. Astron. J.}\ }\textbf {\bibinfo {volume} {190}},\
  \bibinfo {pages} {41} (\bibinfo {year} {2015})},\ \Eprint
  {http://arxiv.org/abs/1407.1504} {arXiv:1407.1504 [gr-qc]} \BibitemShut
  {NoStop}%
\bibitem [{\citenamefont {Moderski}\ and\ \citenamefont
  {Rogatko}(2001)}]{PhysRevD.63.084014}%
  \BibitemOpen
  \bibfield  {author} {\bibinfo {author} {\bibfnamefont {R.}~\bibnamefont
  {Moderski}}\ and\ \bibinfo {author} {\bibfnamefont {M.}~\bibnamefont
  {Rogatko}},\ }\href {\doibase 10.1103/PhysRevD.63.084014} {\bibfield
  {journal} {\bibinfo  {journal} {Phys. Rev. D}\ }\textbf {\bibinfo {volume}
  {63}},\ \bibinfo {pages} {084014} (\bibinfo {year} {2001})}\BibitemShut
  {NoStop}%
\bibitem [{\citenamefont {Zhang}(2017)}]{Zhang:2017zba}%
  \BibitemOpen
  \bibfield  {author} {\bibinfo {author} {\bibfnamefont {H.}~\bibnamefont
  {Zhang}},\ }\href {\doibase 10.1038/s41598-017-03854-y} {\bibfield  {journal}
  {\bibinfo  {journal} {Sci. Rep.}\ }\textbf {\bibinfo {volume} {7}},\ \bibinfo
  {pages} {4000} (\bibinfo {year} {2017})},\ \Eprint
  {http://arxiv.org/abs/1706.08850} {arXiv:1706.08850 [gr-qc]} \BibitemShut
  {NoStop}%
\bibitem [{\citenamefont {Bhadra}(2003)}]{Bhadra:2003zs}%
  \BibitemOpen
  \bibfield  {author} {\bibinfo {author} {\bibfnamefont {A.}~\bibnamefont
  {Bhadra}},\ }\href {\doibase 10.1103/PhysRevD.67.103009} {\bibfield
  {journal} {\bibinfo  {journal} {Phys. Rev. D}\ }\textbf {\bibinfo {volume}
  {67}},\ \bibinfo {pages} {103009} (\bibinfo {year} {2003})},\ \Eprint
  {http://arxiv.org/abs/gr-qc/0306016} {arXiv:gr-qc/0306016} \BibitemShut
  {NoStop}%
\bibitem [{\citenamefont {Karimov}\ \emph {et~al.}(2018)\citenamefont
  {Karimov}, \citenamefont {Izmailov}, \citenamefont {Bhattacharya},\ and\
  \citenamefont {Nandi}}]{Karimov:2018whx}%
  \BibitemOpen
  \bibfield  {author} {\bibinfo {author} {\bibfnamefont {R.}~\bibnamefont
  {Karimov}}, \bibinfo {author} {\bibfnamefont {R.}~\bibnamefont {Izmailov}},
  \bibinfo {author} {\bibfnamefont {A.}~\bibnamefont {Bhattacharya}}, \ and\
  \bibinfo {author} {\bibfnamefont {K.}~\bibnamefont {Nandi}},\ }\href
  {\doibase 10.1140/epjc/s10052-018-6270-6} {\bibfield  {journal} {\bibinfo
  {journal} {Eur. Phys. J. C}\ }\textbf {\bibinfo {volume} {78}},\ \bibinfo
  {pages} {788} (\bibinfo {year} {2018})},\ \Eprint
  {http://arxiv.org/abs/2002.00589} {arXiv:2002.00589 [gr-qc]} \BibitemShut
  {NoStop}%
\bibitem [{\citenamefont {Li}(2013)}]{Li:2013jna}%
  \BibitemOpen
  \bibfield  {author} {\bibinfo {author} {\bibfnamefont {R.}~\bibnamefont
  {Li}},\ }\href {\doibase 10.1103/PhysRevD.88.127901} {\bibfield  {journal}
  {\bibinfo  {journal} {Phys. Rev. D}\ }\textbf {\bibinfo {volume} {88}},\
  \bibinfo {pages} {127901} (\bibinfo {year} {2013})},\ \Eprint
  {http://arxiv.org/abs/1310.3587} {arXiv:1310.3587 [gr-qc]} \BibitemShut
  {NoStop}%
\bibitem [{\citenamefont {Li}\ and\ \citenamefont {Zhao}(2014)}]{Li:2014xxa}%
  \BibitemOpen
  \bibfield  {author} {\bibinfo {author} {\bibfnamefont {R.}~\bibnamefont
  {Li}}\ and\ \bibinfo {author} {\bibfnamefont {J.}~\bibnamefont {Zhao}},\
  }\href {\doibase 10.1140/epjc/s10052-014-3051-8} {\bibfield  {journal}
  {\bibinfo  {journal} {Eur. Phys. J. C}\ }\textbf {\bibinfo {volume} {74}},\
  \bibinfo {pages} {3051} (\bibinfo {year} {2014})},\ \Eprint
  {http://arxiv.org/abs/1403.7279} {arXiv:1403.7279 [gr-qc]} \BibitemShut
  {NoStop}%
\bibitem [{\citenamefont {Huang}\ \emph {et~al.}(2017)\citenamefont {Huang},
  \citenamefont {Liu}, \citenamefont {Zhai},\ and\ \citenamefont
  {Li}}]{Huang:2017whw}%
  \BibitemOpen
  \bibfield  {author} {\bibinfo {author} {\bibfnamefont {Y.}~\bibnamefont
  {Huang}}, \bibinfo {author} {\bibfnamefont {D.-J.}\ \bibnamefont {Liu}},
  \bibinfo {author} {\bibfnamefont {X.-H.}\ \bibnamefont {Zhai}}, \ and\
  \bibinfo {author} {\bibfnamefont {X.-Z.}\ \bibnamefont {Li}},\ }\href
  {\doibase 10.1088/1361-6382/aa7964} {\bibfield  {journal} {\bibinfo
  {journal} {Class. Quant. Grav.}\ }\textbf {\bibinfo {volume} {34}},\ \bibinfo
  {pages} {155002} (\bibinfo {year} {2017})},\ \Eprint
  {http://arxiv.org/abs/1706.04441} {arXiv:1706.04441 [gr-qc]} \BibitemShut
  {NoStop}%
\bibitem [{\citenamefont {Bernard}(2017)}]{Bernard:2017rcw}%
  \BibitemOpen
  \bibfield  {author} {\bibinfo {author} {\bibfnamefont {C.}~\bibnamefont
  {Bernard}},\ }\href {\doibase 10.1103/PhysRevD.96.105025} {\bibfield
  {journal} {\bibinfo  {journal} {Phys. Rev. D}\ }\textbf {\bibinfo {volume}
  {96}},\ \bibinfo {pages} {105025} (\bibinfo {year} {2017})},\ \Eprint
  {http://arxiv.org/abs/1711.05419} {arXiv:1711.05419 [gr-qc]} \BibitemShut
  {NoStop}%
\bibitem [{\citenamefont {Bernard}(2016)}]{Bernard:2016wqo}%
  \BibitemOpen
  \bibfield  {author} {\bibinfo {author} {\bibfnamefont {C.}~\bibnamefont
  {Bernard}},\ }\href {\doibase 10.1103/PhysRevD.94.085007} {\bibfield
  {journal} {\bibinfo  {journal} {Phys. Rev. D}\ }\textbf {\bibinfo {volume}
  {94}},\ \bibinfo {pages} {085007} (\bibinfo {year} {2016})},\ \Eprint
  {http://arxiv.org/abs/1608.05974} {arXiv:1608.05974 [gr-qc]} \BibitemShut
  {NoStop}%
\bibitem [{\citenamefont {Kokkotas}\ \emph {et~al.}(2015)\citenamefont
  {Kokkotas}, \citenamefont {Konoplya},\ and\ \citenamefont
  {Zhidenko}}]{PhysRevD.92.064022}%
  \BibitemOpen
  \bibfield  {author} {\bibinfo {author} {\bibfnamefont {K.~D.}\ \bibnamefont
  {Kokkotas}}, \bibinfo {author} {\bibfnamefont {R.~A.}\ \bibnamefont
  {Konoplya}}, \ and\ \bibinfo {author} {\bibfnamefont {A.}~\bibnamefont
  {Zhidenko}},\ }\href {\doibase 10.1103/PhysRevD.92.064022} {\bibfield
  {journal} {\bibinfo  {journal} {Phys. Rev. D}\ }\textbf {\bibinfo {volume}
  {92}},\ \bibinfo {pages} {064022} (\bibinfo {year} {2015})}\BibitemShut
  {NoStop}%
\bibitem [{\citenamefont {Siahaan}(2015)}]{Siahaan:2015xna}%
  \BibitemOpen
  \bibfield  {author} {\bibinfo {author} {\bibfnamefont {H.~M.}\ \bibnamefont
  {Siahaan}},\ }\href {\doibase 10.1142/S0218271815501023} {\bibfield
  {journal} {\bibinfo  {journal} {Int. J. Mod. Phys. D}\ }\textbf {\bibinfo
  {volume} {24}},\ \bibinfo {pages} {1550102} (\bibinfo {year} {2015})},\
  \Eprint {http://arxiv.org/abs/1506.03957} {arXiv:1506.03957 [hep-th]}
  \BibitemShut {NoStop}%
\bibitem [{\citenamefont {Collins}\ \emph {et~al.}(1973)\citenamefont
  {Collins}, \citenamefont {Delbourgo},\ and\ \citenamefont
  {Williams}}]{Collins:1973xf}%
  \BibitemOpen
  \bibfield  {author} {\bibinfo {author} {\bibfnamefont {P.}~\bibnamefont
  {Collins}}, \bibinfo {author} {\bibfnamefont {R.}~\bibnamefont {Delbourgo}},
  \ and\ \bibinfo {author} {\bibfnamefont {R.}~\bibnamefont {Williams}},\
  }\href {\doibase 10.1088/0305-4470/6/2/007} {\bibfield  {journal} {\bibinfo
  {journal} {J. Phys. A}\ }\textbf {\bibinfo {volume} {6}},\ \bibinfo {pages}
  {161} (\bibinfo {year} {1973})}\BibitemShut {NoStop}%
\bibitem [{\citenamefont {Crispino}\ \emph
  {et~al.}(2009{\natexlab{b}})\citenamefont {Crispino}, \citenamefont {Dolan},\
  and\ \citenamefont {Oliveira}}]{PhysRevD.79.064022}%
  \BibitemOpen
  \bibfield  {author} {\bibinfo {author} {\bibfnamefont {L.~C.~B.}\
  \bibnamefont {Crispino}}, \bibinfo {author} {\bibfnamefont {S.~R.}\
  \bibnamefont {Dolan}}, \ and\ \bibinfo {author} {\bibfnamefont {E.~S.}\
  \bibnamefont {Oliveira}},\ }\href {\doibase 10.1103/PhysRevD.79.064022}
  {\bibfield  {journal} {\bibinfo  {journal} {Phys. Rev. D}\ }\textbf {\bibinfo
  {volume} {79}},\ \bibinfo {pages} {064022} (\bibinfo {year}
  {2009}{\natexlab{b}})}\BibitemShut {NoStop}%
\bibitem [{\citenamefont {Zhang}\ \emph {et~al.}(2013)\citenamefont {Zhang},
  \citenamefont {Berti},\ and\ \citenamefont {Cardoso}}]{PhysRevD.88.044018}%
  \BibitemOpen
  \bibfield  {author} {\bibinfo {author} {\bibfnamefont {Z.}~\bibnamefont
  {Zhang}}, \bibinfo {author} {\bibfnamefont {E.}~\bibnamefont {Berti}}, \ and\
  \bibinfo {author} {\bibfnamefont {V.}~\bibnamefont {Cardoso}},\ }\href
  {\doibase 10.1103/PhysRevD.88.044018} {\bibfield  {journal} {\bibinfo
  {journal} {Phys. Rev. D}\ }\textbf {\bibinfo {volume} {88}},\ \bibinfo
  {pages} {044018} (\bibinfo {year} {2013})}\BibitemShut {NoStop}%
\bibitem [{\citenamefont {Mano}\ \emph
  {et~al.}(1996{\natexlab{a}})\citenamefont {Mano}, \citenamefont {Suzuki},\
  and\ \citenamefont {Takasugi}}]{Mano:1996vt}%
  \BibitemOpen
  \bibfield  {author} {\bibinfo {author} {\bibfnamefont {S.}~\bibnamefont
  {Mano}}, \bibinfo {author} {\bibfnamefont {H.}~\bibnamefont {Suzuki}}, \ and\
  \bibinfo {author} {\bibfnamefont {E.}~\bibnamefont {Takasugi}},\ }\href
  {\doibase 10.1143/PTP.95.1079} {\bibfield  {journal} {\bibinfo  {journal}
  {Prog. Theor. Phys.}\ }\textbf {\bibinfo {volume} {95}},\ \bibinfo {pages}
  {1079} (\bibinfo {year} {1996}{\natexlab{a}})},\ \Eprint
  {http://arxiv.org/abs/gr-qc/9603020} {arXiv:gr-qc/9603020} \BibitemShut
  {NoStop}%
\bibitem [{\citenamefont {Mano}\ and\ \citenamefont
  {Takasugi}(1997)}]{Mano:1996gn}%
  \BibitemOpen
  \bibfield  {author} {\bibinfo {author} {\bibfnamefont {S.}~\bibnamefont
  {Mano}}\ and\ \bibinfo {author} {\bibfnamefont {E.}~\bibnamefont
  {Takasugi}},\ }\href {\doibase 10.1143/PTP.97.213} {\bibfield  {journal}
  {\bibinfo  {journal} {Prog. Theor. Phys.}\ }\textbf {\bibinfo {volume}
  {97}},\ \bibinfo {pages} {213} (\bibinfo {year} {1997})},\ \Eprint
  {http://arxiv.org/abs/gr-qc/9611014} {arXiv:gr-qc/9611014} \BibitemShut
  {NoStop}%
\bibitem [{\citenamefont {Mano}\ \emph
  {et~al.}(1996{\natexlab{b}})\citenamefont {Mano}, \citenamefont {Suzuki},\
  and\ \citenamefont {Takasugi}}]{Mano:1996mf}%
  \BibitemOpen
  \bibfield  {author} {\bibinfo {author} {\bibfnamefont {S.}~\bibnamefont
  {Mano}}, \bibinfo {author} {\bibfnamefont {H.}~\bibnamefont {Suzuki}}, \ and\
  \bibinfo {author} {\bibfnamefont {E.}~\bibnamefont {Takasugi}},\ }\href
  {\doibase 10.1143/PTP.96.549} {\bibfield  {journal} {\bibinfo  {journal}
  {Prog. Theor. Phys.}\ }\textbf {\bibinfo {volume} {96}},\ \bibinfo {pages}
  {549} (\bibinfo {year} {1996}{\natexlab{b}})},\ \Eprint
  {http://arxiv.org/abs/gr-qc/9605057} {arXiv:gr-qc/9605057} \BibitemShut
  {NoStop}%
\bibitem [{\citenamefont {Sasaki}\ and\ \citenamefont
  {Tagoshi}(2003)}]{Sasaki:2003xr}%
  \BibitemOpen
  \bibfield  {author} {\bibinfo {author} {\bibfnamefont {M.}~\bibnamefont
  {Sasaki}}\ and\ \bibinfo {author} {\bibfnamefont {H.}~\bibnamefont
  {Tagoshi}},\ }\href {\doibase 10.12942/lrr-2003-6} {\bibfield  {journal}
  {\bibinfo  {journal} {Living Rev. Rel.}\ }\textbf {\bibinfo {volume} {6}},\
  \bibinfo {pages} {6} (\bibinfo {year} {2003})},\ \Eprint
  {http://arxiv.org/abs/gr-qc/0306120} {arXiv:gr-qc/0306120} \BibitemShut
  {NoStop}%
\bibitem [{\citenamefont {Peters}(1976)}]{PhysRevD.13.775}%
  \BibitemOpen
  \bibfield  {author} {\bibinfo {author} {\bibfnamefont {P.~C.}\ \bibnamefont
  {Peters}},\ }\href {\doibase 10.1103/PhysRevD.13.775} {\bibfield  {journal}
  {\bibinfo  {journal} {Phys. Rev. D}\ }\textbf {\bibinfo {volume} {13}},\
  \bibinfo {pages} {775} (\bibinfo {year} {1976})}\BibitemShut {NoStop}%
\bibitem [{\citenamefont {Lin}\ \emph {et~al.}(2020{\natexlab{b}})\citenamefont
  {Lin}, \citenamefont {Qian}, \citenamefont {Fan},\ and\ \citenamefont
  {Zhang}}]{Lin:2019qyx}%
  \BibitemOpen
  \bibfield  {author} {\bibinfo {author} {\bibfnamefont {K.}~\bibnamefont
  {Lin}}, \bibinfo {author} {\bibfnamefont {W.-L.}\ \bibnamefont {Qian}},
  \bibinfo {author} {\bibfnamefont {X.}~\bibnamefont {Fan}}, \ and\ \bibinfo
  {author} {\bibfnamefont {H.}~\bibnamefont {Zhang}},\ }\href {\doibase
  10.1088/1674-1137/44/7/071001} {\bibfield  {journal} {\bibinfo  {journal}
  {Chin. Phys. C}\ }\textbf {\bibinfo {volume} {44}},\ \bibinfo {pages}
  {071001} (\bibinfo {year} {2020}{\natexlab{b}})},\ \Eprint
  {http://arxiv.org/abs/1903.09039} {arXiv:1903.09039 [gr-qc]} \BibitemShut
  {NoStop}%
\bibitem [{\citenamefont {Hongsheng}\ and\ \citenamefont
  {Xilong}(2018)}]{Hongsheng:2018ibg}%
  \BibitemOpen
  \bibfield  {author} {\bibinfo {author} {\bibfnamefont {Z.}~\bibnamefont
  {Hongsheng}}\ and\ \bibinfo {author} {\bibfnamefont {F.}~\bibnamefont
  {Xilong}},\ }\href@noop {} {\  (\bibinfo {year} {2018})},\ \Eprint
  {http://arxiv.org/abs/1809.06511} {arXiv:1809.06511 [gr-qc]} \BibitemShut
  {NoStop}%
\bibitem [{\citenamefont {Leaver}(1986)}]{PhysRevD.34.384}%
  \BibitemOpen
  \bibfield  {author} {\bibinfo {author} {\bibfnamefont {E.~W.}\ \bibnamefont
  {Leaver}},\ }\href {\doibase 10.1103/PhysRevD.34.384} {\bibfield  {journal}
  {\bibinfo  {journal} {Phys. Rev. D}\ }\textbf {\bibinfo {volume} {34}},\
  \bibinfo {pages} {384} (\bibinfo {year} {1986})}\BibitemShut {NoStop}%
\end{thebibliography}%
\end{document}